\documentclass[fleqn,usenatbib]{mnras}

\usepackage{newtxtext,newtxmath}
\usepackage[T1]{fontenc}

\DeclareRobustCommand{\VAN}[3]{#2}
\let\VANthebibliography\thebibliography
\def\thebibliography{\DeclareRobustCommand{\VAN}[3]{##3}\VANthebibliography}

\usepackage{graphicx}	% Including figure files
\usepackage{amsmath}	% Advanced maths commands
\usepackage{subfig,graphicx}
\usepackage{longtable}
\usepackage{supertabular}
\usepackage{lscape}
\newcommand{\cross}[1][1pt]{\ooalign{%
  \rule[1ex]{1ex}{#1}\cr% Horizontal bar
  \hss\rule{#1}{.7em}\hss\cr}}% Vertical bar

\def \kms {km~$\rm s^{-1}$ }

\title[Gas depletion in luminous quasars at $z\sim2$]{Evidence for universal gas depletion in a sample of 41 luminous Type 1 quasars at $z\sim2$}

\author[S. J. Molyneux et al.]{
S. J. Molyneux$^{1}$\thanks{Email: smolyneux.astro@gmail.com},
M. Banerji$^{1}$, M. J. Temple$^{2}$, M. Aravena$^{3}$, R. J. Assef$^{3}$, P. Hewett$^{4}$, G. C. Jones$^{5, 6}$, \newauthor A. Puglisi$^{1}$, A. L. Rankine$^{7}$, C. Ricci$^{3, 8}$, M. Stepney$^{1}$, S. Tang$^{1}$ 
\\
$^{1}$School of Physics and Astronomy, University of Southampton, Southampton SO17 1BJ, UK\\
$^{2}$Centre for Extragalactic Astronomy, Department of Physics, Durham University, South Road, Durham DH1 3LE, UK\\
$^{3}$ Instituto de Estudios Astrofísicos, Facultad de Ingeniería y Ciencias, Universidad Diego Portales, Avenida Ejercito Libertador 441, Santiago, Chile\\
$^{4}$ Institute of Astronomy, University of Cambridge, Madingley Road, Cambridge CB3 0HA, UK \\
$^{5}$ Kavli Institute for Cosmology, University of Cambridge, Madingley Road, Cambridge CB3 0HA, UK\\
$^{6}$ Cavendish Laboratory, University of Cambridge, 19 JJ Thomson Avenue, Cambridge CB3 0HE, UK \\
$^{7}$ Institute for Astronomy, University of Edinburgh, Royal Observatory, Edinburgh EH9 3HJ, UK\\
$^{8}$ Kavli Institute for Astronomy and Astrophysics, Peking University, Beijing 100871, China
}
\date{Accepted XXX. Received YYY; in original form ZZZ}

\pubyear{2025}

\begin{document}
\label{firstpage}
\pagerange{\pageref{firstpage}--\pageref{lastpage}}
\maketitle

\begin{abstract}
We present ALMA CO observations of the molecular gas in a sample of 41 luminous unobscured quasars at z $\sim$ 2 from the Sloan Digital Sky Survey. 32 targets comprise the main sample observed in CO(3-2) and 9 targets have archival ALMA data of CO(3-2), CO(4-3) and CO(7-6). All quasars have rest-UV to optical spectra tracing ionised gas in the broad line region (e.g. C~\textsc{iv}) and the narrow line region (e.g. [O~{\sc iii}]) and they cover the full range of outflow properties in the SDSS quasar population at these redshifts. 15 out of the 32 quasars in the main sample are detected in CO(3-2) and five out of the nine archival quasars are also detected in CO. The median gas mass for all 20 CO detected quasars is 8.0~$\pm$~1.5~$\times$~10$^9$~M$_{\odot}$ with a median M$_{\rm dyn}$ of 1.4~$\pm$~0.9~$\times$~10$^{11}$~M$_{\odot}$. We find gas fractions in the range 0.02 – 0.32, which are generally lower than both inactive galaxies and obscured quasars at similar redshifts. We suggest an evolutionary trend in gas fractions of quasar host galaxies from obscured and gas rich to unobscured and gas poor. We note a tentative correlation between the gas fractions and the broad-line region properties with quasars showing high C~\textsc{iv} blueshifts, indicating stronger broad-line region winds, having higher gas fractions. Six of the quasars corresponding to 15\% of the sample also show evidence for at least one companion galaxy detected in CO at the same redshift.

\end{abstract}


%%%%%%%%%%%%%%%%%%%%%%%%%%%%%%%%%%%%%%%%%%%%%%%%%%

%%%%%%%%%%%%%%%%% BODY OF PAPER %%%%%%%%%%%%%%%%%%

\section{Introduction}
\label{sec:intro}

%\com{Broad picture, galaxy BH co-evolution. CO key ingredient, traces molecular gas, fuel for star formation. Ability of AGN to impact upon galaxy via feedback, radiation, winds, outflows.}

%\com{AGN necessary}
Feedback from active galactic nuclei (AGN) is required to explain our current understanding of galaxy evolution, as it is thought to regulate the observed co-evolution of accreting black holes (BH) and their host galaxies across cosmic time \citep[e.g.][]{Fabian12, Cresci18}. To understand the physical processes by which feedback can impact the host galaxy, we need to characterise the interactions between central supermassive black holes and the host galaxy interstellar medium (ISM) across multiple gas phases (e.g., molecular, ionised, atomic). This requires observations at different wavelengths and spatial scales.

%\com{Outflows observed in ionised and molecular phase}
Studies investigating the properties of the multi-phase ISM in quasar host galaxies find the presence of kpc-scale outflows in the ionised phase traced by [O~{\sc iii}]5007{\AA} \citep[e.g.][]{Harrison14, Carniani15, Cresci15, Circosta18, Scholtz20, Kakkad20, Vayner21, Concas22, Wylezalek22} and also in the molecular phase traced by carbon monoxide (CO) \citep[e.g.][]{Cicone12, Feruglio15, Brusa18, Bischetti19, Longinotti23}. These multi-phase outflows have been seen to correlate with AGN properties such as luminosity \citep[e.g.][]{Mullaney13, Fiore17} and the presence of radio jets \citep[e.g.][]{Molyneux19}. These observations show the ability of AGN activity to expel the star-forming ISM and would suggest that AGN can have an impact on the gas reservoirs of their host galaxies. The molecular phase of the ISM is an important consideration, since this gas is redistributed to promote star formation activity and fuel BH growth \citep[e.g.][]{McKee07, Carilli13, Vito14, Tacconi20}, and therefore plays a critical role in galaxy evolution.

To analyse the molecular phase of the ISM, CO is routinely observed, which can provide an instantaneous measure of the current available fuel for star formation \citep{Bolatto13}. We can trace an impact by AGN activity by observing the gas fractions (the ratio of the gas mass to the stellar mass). Indeed, some luminous quasars at high redshift (z~$>$~2) that are driving powerful winds show depleted gas reservoirs compared to their non-AGN counterparts \citep[e.g.][]{Circosta21, Bischetti21, Bertola24}. However, AGN at lower redshifts (z $<$ 0.3) are also routinely observed to be hosted in gas-rich, star-forming galaxies, with no signs of a depleted molecular gas reservoir compared to non-AGN \citep[e.g.][]{Saintonge17, Shangguan20, Jarvis20, Lamperti20, Koss21}. Simulations also support the picture from the local Universe with AGN having higher gas fractions than non-AGN, and residing in gas-rich, star-forming galaxies \citep{Piotrowska22, Ward22}. These results indicate that powerful AGN/quasars do have the ability to have a significant impact on the global properties of their host galaxies, but that gas-rich environments are also needed to fuel the most powerful quasars. The molecular gas properties of AGN also depend on redshift, luminosity and obscuration \citep[e.g.][]{Perna18, Circosta21, Banerji21, Bischetti21, Sun24}.  %\com{Mention here that z=2 quasars also show high gas fractions in the simulations - Samuels 2022 paper, EAGLE, TNG and SIMBA simulations} 

%To fully understand the observational results, an important factor that needs to be taken into account are the selection effects, which can include the redshift of the sample, the luminosity of the AGN and the amount of dust present (potentially indicating different evolutionary phases). Further the type of AGN (which is also linked to the dust properties) should be considered. For example, dusty and obscured red quasars are found to have similar or lower gas fractions to galaxies on the main-sequence \citep[e.g.][]{Perna18, Circosta21, Banerji21}. At the more extreme end, hot, dust-obscured galaxies (HotDOGs) at z $>$ 2 which host luminous quasars seem to lie closer to the main sequence \citep{Sun24}. In comparison, luminous unobscured blue quasars at z $\sim$ 2 seem to show much lower gas fractions \citep{Bischetti21}. Samples of quasars with analysis of their gas fractions however are limited to a few tens of sources and we therefore require larger samples to analyse differences in these populations.

%Furthermore, timescales are likely to play a key role, since how long feedback processes have been ongoing (e.g. outflows) will affect the time that they have had chance to have an impact on the global galactic properties.

%\com{Further investigate sub-parsec outflows - rest frame UV spectra - CIV}
In addition to multiphase outflows on kpc scales, luminous quasars also show evidence of outflows on sub-parsec scales. High ionisation emission lines such as C~\textsc{iv}1550{\AA} that show strong asymmetries to the blue, can be indicative of radiatively-driven accretion disk winds in the quasar broad line region \citep[BLR e.g.][]{Richards11, Rankine20, Stepney23}. There are established correlations between the prevalence of these strong winds and the ultraviolet (UV) SEDs of quasars \citep[e.g.][]{Vietri20, Temple23}. For example, the He~\textsc{ii} 1640\AA\, emission line provides a probe of the hardness of the ionizing SED \citep{Leighly04, Baskin13, Baskin15} and is also correlated to C~\textsc{iv} blueshift with quasars that have softer ionising SEDs driving stronger outflows \citep{Temple23}. Quasars with and without broad absorption lines (BALs), which are considered a more direct probe of high-velocity winds, also appear to have very similar emission line properties \citep{Rankine20} as well as showing a similar relationship between emission line morphology and ionising SED as the non-BAL population \citep{Temple23}. 

%However, it should be noted that it is still debated as to whether BALs trace gas close to the BH or trace gas along the line of sight that is much further away from the BH \com{CITE}. , as well as broad absorption features in quasar spectra \citep{Rankine20}

 %If the quasar ionising SED is sufficiently soft, electrons can remain bound to the nuclei and radiation line-driving contributes to an acceleration of material which can lead to stronger outflows as measured via the blueshifts of these high-ionization UV emission lines \citep{Rankine20}.

%\com{Correlation identified with [OIII]}
To test whether there is any link between the rest frame UV spectra and the wider, global properties of the host galaxy, we can study correlations between the UV spectra and the global ISM properties. As discussed before, [O~{\sc iii}] is a useful tracer of the ionised phase of the ISM in the narrow-line region (NLR) and therefore a good emission line to use in these analysis. Winds from the accretion disk, traced by the blueshift of C~\textsc{iv}, have been shown to correlate with the velocity of winds in the quasar NLR as traced by [O~{\sc iii}] \citep{Coatman19, Vietri20, Temple24}. Since the [O~{\sc iii}] emission traces outflows on larger scales than C~\textsc{iv}, a correlation in the kinematics of these two ionised gas tracers is consistent with the scenario that quasar disk winds could be responsible for the impact observed on the ISM at kiloparsec scales. An interesting avenue of research is therefore to test whether any such link can be made to the molecular gas (CO) and make comparisons to ionised outflows from the NLR ([O~{\sc iii}]) and the BLR (C~\textsc{iv}). 

In addition to the properties of the ISM host galaxy, there are also unanswered questions relating to the environments of luminous high-z quasars and whether or not they reside in over dense regions. Several previous studies have identified companion galaxies to quasar host galaxies at $z$ = 2 -- 6 either in individual systems \citep{Ivison08, Salome12, Fogasy17, Banerji17, Banerji21, Decarli17, Trakhtenbrot17, Neeleman19, Fogasy20, Stacey22, Li23}. %\cite{} identified 70 per cent of their sample of quasars to have companions, when using deep observations of CO(4-3).
The samples studied are however small, which motivates work across larger samples to determine whether over-densities are prevalent across all luminous AGN and quasars.

In this work we present a comprehensive study of the molecular gas properties of 41 quasars at $z\sim$~2, traced by CO observations with ALMA. The sample has already been well studied, with analysis of the rest frame UV spectra \citep{Rankine20, Temple23} and rest-frame optical spectra \citep{Temple24}. These detailed spectroscopic characterisations make the sample unique, with information about the ionised gas across different spatial scales. We can now explore the relationship between the sub-parsec scale nuclear region (traced by the rest-frame UV spectra) and NLR ([O~{\sc iii}]) to the global molecular ISM - traced by CO. We compare our results to samples of star forming galaxies, AGN and quasars across the redshift range 0 -- 5 from the literature.

In Section~\ref{sec:observations} we introduce the sample of quasars and the data used in this work. In Section~\ref{sec:results} we describe the analysis techniques used to study the molecular gas and dust continuum. We calculate gas masses and gas fractions and compare to other populations in the literature to place our findings in context. In Section~\ref{sec:discussion} we discuss our findings and our final conclusions are presented in Section~\ref{sec:conclsions}. 

We adopt a flat LambdaCDM cosmology with parameters: $H_0=70$\,\kms\,Mpc$^{-1}$, $\Omega_M=0.3$, and $\Omega_\Lambda=0.7$ throughout.

\section{Observations and ancillary data}
\label{sec:observations}

\subsection{Sample Selection}
\label{sec:sample_select}

The quasars presented here were selected from a larger parent sample of $\sim$144k quasars from the Sloan Digital Sky Survey Data Release 14 (SDSS DR14) at $1.6 \lesssim z \lesssim 3.5$ where the rest-frame UV spectra have been analysed in detail to constrain their accretion and outflow properties \citep{Rankine20,Temple23}. We restricted the sample to the subset of quasars with rest-frame optical spectra covering [O~{\sc iii}] from either the Gemini GNIRS Distant Quasar Survey \citep{Matthews21} or the compilation of \citet{Coatman19}. The [O~{\sc iii}] data enable us to constrain outflows on larger scales in the quasar NLR, and relate this to the molecular gas properties. To ensure high-quality [O~{\sc iii}] measurements we restrict the parent sample to magnitudes \textit{i}$_\textrm{AB}$~$<$~19.1 and a limited range in redshift of $2.2 \lesssim z \lesssim 2.4$. 

To generate a sample to target with ALMA, we then identified a subset of 50 of these quasars with rest-UV and rest-optical spectroscopy which cover the full range in C~\textsc{iv} blueshifts and equivalent widths as the parent SDSS sample. Due to the known correlations between C~\textsc{iv} and [O~{\sc iii}] emission line properties \citep{Vietri18,Coatman19,Temple24} this also results in a range of [O~{\sc iii}] line widths and equivalent widths. Our sample selection therefore allows us to investigate how the molecular gas properties of quasar host galaxies depend on the outflow properties measured from their rest-frame UV and optical spectra. Of these 50 targets, CO(3-2) was observed in 32 quasars during ALMA Cycle 8, which therefore comprise the main sample of quasars presented here. Our sample includes 21 broad absorption line (BAL) quasars, allowing us to look at differences in molecular gas properties between BALs and non-BALs. The sample deliberately covers a limited range in luminosity and black hole mass [median log$_{10}(L_{3000}$ / erg\,s$^{-1}$) = 46.7 and log$_{10}$($M_{\rm BH}$ / M$_{\odot}$) = 9.48; \citealt{Temple23}], allowing us to explicitly test the dependence of molecular gas properties on outflow properties measured from the rest-UV and optical spectra.
% \textcolor{red}{TO DO: Check that the ones that weren't observed are not biased in any way in terms of their properties}.  

\begin{figure*}
\centering
\includegraphics[width=0.9\textwidth]{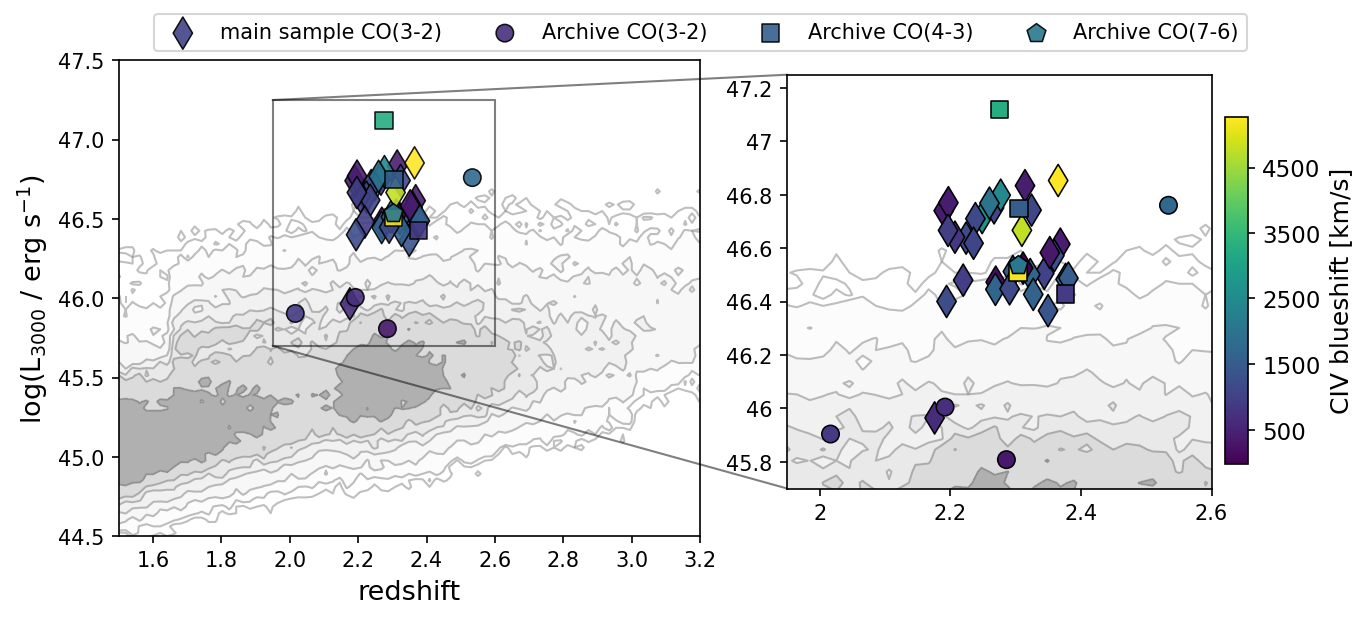}
\caption{Here we present the sample selection for the quasars presented in this work. We plot the sample in the redshift and 3000\AA\, luminosity parameter space, coloured by the C~\textsc{iv} blueshifts. The grey contours/shaded region indicate the full distribution of SDSS DR16 quasars from \citealt{Rankine20, Temple23}. Diamonds indicate the main sample from ALMA project ID 2021.1.00393.S (PI: M. Temple). We also show archival ALMA data utilised within this work, with CO(3-2) data presented as circles, CO(4-3) data presented as squares, and CO(7-6) data presented as hexagons. For a full description of the sample selection, see Section~\ref{sec:sample_select}.
}
\label{fig:sample-select}
\end{figure*}

In addition to the main sample of 32 quasars, we identified another 9 targets using the ALminer tool \citep{Ahmadi23} to match all quasars from the parent \citet{Rankine20} sample with the ALMA archival database, using a matching radius of 0.75 arcmin (corresponding to $\sim$ 360\,kpc, the primary beam of ALMA for the main sample presented here). These archival data have the same selection criteria as the main sample and are included to increase the sample size. These additional targets are referred to as the archival sample and include observations of the CO(3-2) emission of 4 targets [project IDs,  2013.1.01262.S (PI: J. Prochaska), 2016.1.00798.S (PI: V. Mainieri), 2017.1.01676.S (PI: C. Ross)], CO(4-3) emission of 4 targets [project ID 2019.1.01251.S (PI: B. Emonts) \citep[see][]{Li23}], and CO(7-6) in 1 target [project 2018.1.00583.S (PI. F. Hamann)]. The redshift and luminosity for the sample colour coded by their C~\textsc{iv} blueshift is presented in Fig.~\ref{fig:sample-select}.

\subsection{ALMA observations}
\label{sec:alma_obs}

We used 12-m array ALMA band 3 to observe the molecular gas emission traced by the CO(3–2) transition in a sample of 32 luminous quasars at $2.2 < z < 2.4$. These observations were taken in 2022 January -- September, observed under the project ID 2021.1.00393.S (PI: M. Temple), in the configurations C43-1, C43-2, C43-3 and C43-4. We refer to these targets as the main sample of this work. Further details of the observations can be found in Table \ref{tab:obs_props}. 

We observed the CO(3-2) emission line down to a uniform sensitivity limit of $\sim$~0.5mJy per 33~\kms bin width (see Table~\ref{tab:obs_props}) and at a resolution of 1 -- 3 arcsec, to obtain unresolved observations of the total molecular gas content. The spectral windows of the observations were aligned to cover the CO(3-2) with one window centred at the expected CO(3-2) frequency based on the systemic redshifts calculated for these quasars as described in Section 3 of \citet{Rankine20}. Briefly, the redshifts are derived from spectral reconstructions based on a Mean-Field Independent Component Analysis (MFICA) technique applied to spectra with an improved sky-subtraction routine relative to the SDSS pipeline. The redshift estimation routine uses the rest-frame 1600-3000\AA\, region, which deliberately excludes the C~\textsc{iv} emission line which can be significantly blueshifted relative to systemic. This is a key difference relative to the SDSS pipeline redshifts and should result in more robust systemic redshifts from the UV spectra. More details of the UV redshift estimates can be found in \citet{Rankine20}. 

As shown in Table~\ref{tab:obs_props}, the archival samples (introduced in Section~\ref{sec:sample_select}) reach different sensitivity limits and some also trace %as well as in some cases tracing
different CO transitions compared to the main sample. When analysing correlations between the molecular gas and the UV/optical line properties (Section~\ref{sec:Correlations}), we therefore only use the main sample, which has a uniform sensitivity.

%The archival sample (introduced in Section~\ref{sec:sample_select}) reach different sensitivity limits \com{and some also trace} %as well as in some cases tracing
%different CO transitions, which needs to be taken into account in the analysis presented in this paper. 
%For this reason, \com{but exclude them when analysing correlations with the UV/optical line properties.}
%we only include the archival sample when analysing the gas fractions,

%%%%%%Manda%%%%

%we choose to present the CO spectra for those data taken from the archive and use these in our analysis of gas fractions, however for analysis of correlations with the UV/optical line properties, we limit this to those from the main sample.

All the data utilised here are the reduced ALMA data products available from the ALMA archive which have been continuum subtracted and primary beam corrected. %For the main sample the \textsc{tclean} parameters used to create the data files were as follows: Briggs weighting with robust = 0.5, a clean threshold of 0.667mJy and a cell size of 0.19 arcsec. 
To make sure these products were reliable, we carried out our own reductions of some of the targets within the sample, namely J2352$-$0120, J1113+1022 and J2239$-$0047 which covered the full range of beam sizes (1 -- 3 arcsec), allowing us to test whether sources had been resolved. %J2352$-$0120 in particular was found to be slightly spatially resolved as well as having a double peaked CO profile and therefore served as a good test. 
We used the \textsc{tclean} function in \textsc{casa} to test different reductions with different parameters such as the weighting (natural and briggs with robust -0.5, 0 and 0.5) as well as the cleaning thresholds ranging from 0.5 to 2 times the threshold used to create the original ALMA product. Quantitatively, we identified a variation in the total line flux obtained on the order of $<$ 2 per cent, well within the uncertainties measured.

\subsection{Ancillary data}
\label{sec:ancillary_data}

In addition to ALMA observations of CO, all 41 targets possess ancillary multi-wavelength data as described below.

\subsubsection{Rest frame UV data}

For all targets presented in this paper we have rest frame UV spectroscopy obtained from SDSS DR14 \citep{Paris18}. For the purposes of this paper we make use of the MFICA reconstructions produced by \citet{Rankine20}, who use the properties of the 1900\,\AA\ emission-line blend to place priors on the MFICA component weights, allowing the intrinsic emission-line profile of C~\textsc{iv}\,$\lambda1549$ to be reconstructed even in objects with broad absorption features. For this paper we use the equivalent width (EW) of He~\textsc{ii}\,$\lambda1640$ and the emission-line blueshift of C~\textsc{iv} as tracers of the high-ionisation broad line region. As discussed at length in \citet{Temple23}, the He~\textsc{ii} EW traces the strength of the unseen ionising continuum at 54\,eV, while the C~\textsc{iv} blueshift provides a measure of outflowing disk-winds from the BLR. We assume here that the blueshifted C~\textsc{iv} emission line profile is tracing an outflow along the line-of-sight \citep{Leighly04, Richards11}. Together these emission lines trace the sub-parsec scale outflow and ionisation properties, which can be correlated with the global ISM properties inferred from the ALMA data. 
% Values of C~\textsc{iv} blueshift range from 10 -- 5000 \kms. \manda{more info needed here}

Using $L_{3000}$ from our rest frame UV data we can calculate a bolometric luminosity using a correction factor determined by the following equation from \cite{Netzer19}:

\begin{equation}
k_{\rm BOL} = 25 \times (L_{3000}/10^{42})^{-0.2}
\end{equation}

For our sample we find correction factors in the range 2.6 -- 4.3, and applying these gives us bolometric luminosities in the range 10$^{46.4-47.5}$ erg s$^{-1}$.

\subsubsection{Near-infrared spectra and BH masses}

Every quasar in our main sample has a high-quality near-infrared spectrum from either \citet{Coatman19} or the GNIRS Distant Quasar Survey \citep{Matthews21}. These spectra were modelled homogeneously by \citet{Temple24} to quantify emission-line properties of [O~{\sc iii}] and H\,$\beta$. Each spectrum was modelled using \texttt{Fantasy} \citep{Ilic23} to include a power-law continuum, Fe~{\sc ii} multiplets, two Gaussian components for the broad Balmer lines, and two Gaussian components for each of the [O~{\sc iii}]\,$\lambda\lambda4960,5008$ lines. From these models we measure $W_{80}$, the 80 per cent velocity width of [O~{\sc iii}], which is a measure of the strength of ionised gas outflows in the narrow-line region. For the objects considered here $W_{80}$ lies in the range 1000 -- 2400 \kms.

Using the full-width at half-maximum (FWHM) of the broad Balmer H\,$\beta$ emission line, and $L_{5100}$ we estimate the black hole mass of each object using the following single-epoch scaling relation derived by \citet{Shen24}:

\begin{equation}
    log \Bigg(\frac{M_{\rm BH, H\beta}}{M_\odot}\Bigg) = log \Bigg[\Bigg(\frac{L_{5100,~\rm AGN}}{10^{44}~\rm erg s^{-1}} \Bigg)^{0.5} \Bigg(\frac{FWHM}{\rm km s^{-1}} \Bigg)^2  \Bigg] + 0.85.
\end{equation}

\begin{table*}
\centering
\begin{tabular}{ |c|c|c|c|c|c| } 
 \hline
 \rule{0pt}{2ex}
 Line & Source Name & Beam Size & Line RMS, $\Delta v$ = 50 \kms & $\rm t_{obs}$  & continuum rms  \\
  &  & (arcsec) & (mJy) & (minutes) & ($\mu$Jy) \\
 \hline
 \rule{0pt}{2ex}
   CO(3-2) main sample & J0014+0912 & 1.124 & 0.31 & 55.4 &  13.8 \\
   ALMA project id: 2021.1.00393.S & J0019+1555 & 2.876  & 0.35 & 30.2 & 14.0  \\
   & J0104+1010 &  1.162 & 0.42 &  27.7 & 13.8    \\
   & J0105+1942 &  1.234 & 0.50 & 30.7  & 13.1    \\
   & J0106+1010 & 1.381 & 0.27 &  55.4  & 14.2    \\
   & J0106$-$0315 & 1.016 & 0.53 & 27.2  &  14.0   \\
   & J0140$-$0138 & 1.014 & 0.34 &  27.2  & 14.0    \\
   & J0142+0257 & 1.794 & 0.28 & 27.2  &  14.2   \\
   & J0351$-$0613 & 2.648 & 0.27 & 27.2 & 13.8    \\
   & J0758+1357 & 1.785 & 0.24 & 62.5 & 15.0    \\
   & J0810+1209 & 2.892 & 0.39 & 20.2  & 16.7    \\
   & J0811+1720 & 1.759 & 0.35 & 30.2  & 16.8    \\
   & J0815+1540 & 1.190 & 0.28 & 30.7 &  12.2   \\
   & J0826+1434 & 2.938 & 0.34 & 29.2  & 14.1   \\
   & J0826+1635 & 1.186 & 0.30 & 32.8  & 14.6    \\
   & J0827+0618 & 1.209 & 0.37 & 29.7  & 14.5    \\
   & J0832+1823 & 1.179 & 0.43 &  30.7 & 12.1    \\
   & J0837+0521 & 1.175 & 0.36 & 27.2  & 15.0    \\
   & J1113+1022 & 3.086 & 0.26 & 28.7  & 13.4    \\
   & J1213+0807 & 1.073 & 0.33 & 27.2  & 13.7    \\
   & J1251+1143 & 1.030 & 0.42 & 30.7  & 13.5    \\
   & J1532+1739 & 1.168 & 0.32 & 29.7  & 13.8    \\
   & J1606+1735 & 1.161 & 0.44 & 30.7  & 13.9    \\
   & J2059$-$0643 & 2.560 & 0.24 & 26.2  & 14.3    \\
   & J2108$-$0630 & 1.254 & 0.63 & 25.7  & 15.7    \\
   & J2239$-$0047 & 2.399 & 0.38 & 27.7  & 14.1    \\
   & J2256+0105 & 2.561 & 0.37 & 27.2  & 14.6    \\
   & J2256+0923 & 2.695 & 0.35 & 28.2  & 14.5    \\
   & J2300+0031 & 0.992 & 0.31 & 29.2  & 13.9    \\
   & J2314+1824 & 2.844 & 0.27 & 30.7  & 13.1    \\
   & J2348+1933 & 2.805 & 0.36 & 33.8  & 14.1    \\
   & J2352$-$0120 & 1.075 & 0.29 & 26.2  & 14.3    \\
  \hline 
  \rule{0pt}{2ex}
  CO(3-2) archive   & J0229$-$0402 & 0.974 & 0.39 & 9.576 &  20.0 \\
   ALMA project id: 2016.1.00798.S & J1000+0206 & 0.660 & 0.33 & 9.072 &  17.6  \\
   \hline 
  \rule{0pt}{2ex}
  CO(3-2) archive & J1420+1603 & 0.520 & 0.26 & 38.3 &  11.5 \\
 ALMA project id: 2013.1.01262.S &  &  &  &  &     \\
  \hline 
  \rule{0pt}{2ex}
    CO(3-2) archive & J1625+2646 & 1.941 & 0.28 & 15.6 & 17.4  \\
 ALMA project id: 2017.1.01676.S &  &  &  &  &     \\
  \hline 
  \rule{0pt}{2ex}
  CO(4-3) archive &  J0052+0140 & 1.859 & 0.18 & 63.5 & 12.6  \\ 
    ALMA project id: 2019.1.01251.S & J1416+2649 & 2.107 & 0.21 & 143.1 & 12.2    \\
    & J2121+0052  & 1.247 & 0.30 & 64.5 &  11.7   \\
    & J2123$-$0050  & 1.822 & 0.18 & 62.5 &  13.8   \\
   \hline 
  \rule{0pt}{2ex}
   CO(7-6) archive  & J1006+0119 & 0.765 & 0.33 & 48.4 & 13.7  \\
   ALMA project id: 2018.1.00583.S &  & &  &    \\
  \hline 
 \end{tabular}
\caption{Details of the CO observations including the observed CO line and ALMA project id, the source name, beam size, line rms, observing times ($\rm t_{obs}$) and continuum rms.}
\label{tab:obs_props}
\end{table*}

\section{Analysis and results}
\label{sec:results}

\subsection{Spectral fitting}
\label{sec:spectral_fitting}

To extract the CO spectra for each quasar, we  use an iterative process of analysing the emission at the expected location of the quasar and identify the spectral region that encompasses the entire line emission, from which we create a narrowband image.
%created a narrowband image containing the emission line in the central pixel of the cube. From the resulting narrowband we select the brightest pixel and again create a narrowband that encompasses the spectra in this pixel. We repeat the process until a consistent narrowband surrounding the spectra is reached. 
The only exceptions for this was J1006+0119 where the detected emission was not centred in the cube (the expected location of the source). For this target we manually searched the region around the expected location and identified the emission slightly offset from the centre of the cube.

From the resulting collapsed images we derived the size of the emission by fitting the images using the \textsc{imfit} routine in \textsc{casa}. The emission size, described by the major and minor axis as well as the position angle, was then used as the aperture to extract the final spectra. To confirm these apertures captured the total flux, we examined the total line intensity with an aperture of increasing radii starting from the sizes given by the \textsc{imfit} routine until the line intensity no longer increased. We confirm that in all but 2 cases, the full emission was captured. The 2 exceptions are J0014+0912 and J1251+0807 where the line intensity increased at increasing radii beyond the initial \textsc{imfit} aperture. Possible reasons for this could be extended diffuse emission and/or companions that are not bright enough to be confirmed. For these 2 sources we therefore take the aperture at which the line intensity flattens off and remains consistent with increasing aperture size.

%For those without any visible emission we extracted the spectra using the beam size centred on the target coordinates as an aperture. From the extracted spectra from these regions we determined the signal-to-noise-ratio ratio of any emission line and kept only those with S/N $>$ 3. The S/N was calculated by dividing the total line intensity by the rms intensity.

\begin{figure}
\centering
\includegraphics[width=\columnwidth]{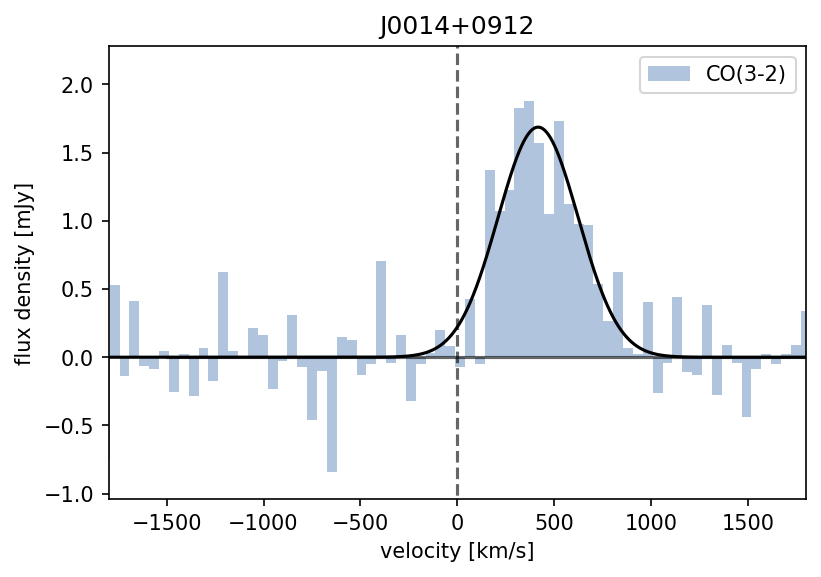}
\caption{Example spectrum and spectral fitting for the target J0014+0912. We present the flux density versus velocity, in velocity bins of 50 \kms. The solid black line shows the Gaussian fit to the emission line and the dashed vertical line represents the zero velocity at the expected redshift determined from H$\alpha$. The same format is used for all other spectra presented in this work (shown in Fig.~\ref{fig:all_spectra}).
}
\label{fig:example_spectra}
\end{figure}

We performed a rebinning such that all spectra were at $\sim$ 50 \kms resolution, with the exception of CO(7-6) which was binned to $\sim$ 40 \kms resolution. This allowed us to clearly identify the presence of any spectral lines within the data. Spectra (example shown in Fig.~\ref{fig:example_spectra}) that possess an emission line with velocity-integrated S/N $>$ 3 are then analysed using Gaussian profiles. Single and double Gaussians are fitted to each emission line with the double-Gaussian parametrization adopted when the reduced chi-square improves by more than 10 per cent. The resulting line-luminosities and FWHM are presented in Table~\ref{tab:CO_props}.

For non-detections we present 3 sigma upper limits for the integrated line intensity. These are based on calculating the 1 sigma rms from the spectra over line free regions, multiplying this by 3 to reach a 3 sigma upper limit for the peak flux. We then assume a line width that is equal to the median of those with detections and use this to model a Gaussian for which we can calculate the upper limits for the line intensity and line luminosity.

We find a $\sim$50 per cent detection ratio in the main sample of 32 targets which have a range of FWHM$_{\rm CO}$ from 170 -- 620 \kms. %168 -- 619 \kms. 
Integrated line intensities for these detections range from 150 -- 1900 \kms.
%145 -- 1888 mJy \kms. 
Three of these targets are also found to have double peaked profiles. Without spatially resolved data it is unclear as to the origin of the double peaks, which could be the result of rotation, outflows or mergers. For BALs five out of 11 are detected and for non-BALs 14 out of 25 are detected, corresponding to 45 and 56 per cent respectively, but given the sample size this is not statistically significant.

The archival data are reduced and analysed in the same way as the main sample. From the same analysis we find non-detections for all CO(3-2) archival data, which have rms sensitivities comparable with those of our main sample (median of 0.31 mJy compared to the main sample median of 0.345 mJy; see Table~\ref{tab:obs_props}). However, we find detections for all 4 targets with CO(4-3) data and also the one target with CO(7-6), all of which are fitted with a single Gaussian. These have rms sensitivities of 0.195 mJy (median) and 0.33 mJy respectively.

\subsection{CO vs H$\alpha$ and UV redshifts}

\begin{figure}
\centering
\includegraphics[width=\columnwidth]{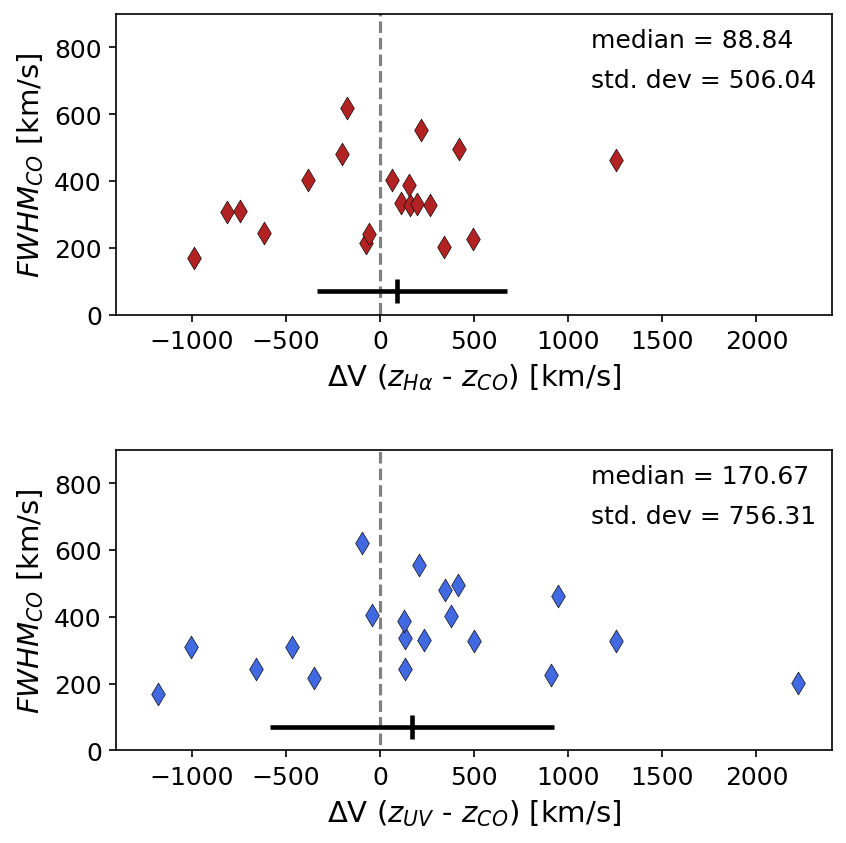}
\caption{Figure showing the velocity offsets between the CO redshift and the H$\alpha$ redshift (top panel) and UV redshift (bottom panel). These are plotted against the FWHM$_{\rm CO}$.
}
\label{fig:v50_vs_COFWHM_balmer_UV}
\end{figure}

%\com{why h alpha not h beta?}
For those quasars with CO detections we were able to measure the CO redshifts and compare them to the redshifts obtained from both the rest-frame optical (from a joint fit to the broad H$\alpha$ and H$\beta$ lines, from NIR data in \citealt{Temple24}) %(derived from the peak of broad H$\alpha$, from NIR data in \citealt{Temple24})
and the rest-frame UV spectra (derived from the SDSS spectrum, as described above in Section~\ref{sec:alma_obs}). [O~{\sc iii}] would provide a better estimate of the redshifts than H$\alpha$, however in the sample several of the quasars were weak or undetected in [O~{\sc iii}] and so H$\alpha$ was used as an alternative to be consistent across the sample. The CO redshifts are determined from the V$_{50}$ of CO line detection (the median velocity of the overall emission-line profile), which are mostly consistent with the H$\alpha$ and UV redshift determinations. The velocity offsets compared to the H$\alpha$ and UV redshifts are presented in Fig.~\ref{fig:v50_vs_COFWHM_balmer_UV} where we present the individual points as well as the distribution of these data, showing that there is a closer match between the H$\alpha$ line velocities and those of the CO(3-2), compared to what is found from the rest frame UV. With a H$\alpha$ measurement, it can be seen that the maximum differences are $\sim$ 1000 \kms, meaning that observations targeting a CO transition based on a H$\alpha$ redshift would still likely remain within the spectral setup with ALMA, despite the scatter. With the UV spectra from SDSS there is more scatter, with up to 2000 \kms offsets found. The majority of our CO redshifts agree well with the UV and optical redshifts, mostly within a few hundred \kms which is the typical uncertainty at these redshifts. The robustness of our rest-UV and rest-optical redshift estimates means we are confident that the CO line would be present in the ALMA spectral window for all non-detections, which therefore allows us to derive meaningful upper limits. 

%There are a few quasars with $>$ 500 \kms offsets between the CO and UV/optical redshift, however there is no observed trend with CO FWHM. For 6 out of 8 targets with offsets greater than 500 \kms differences they are without [O~{\sc iii}] detections, which makes the optical redshifts less robust and may explain the larger observed offsets. 

\subsection{Dynamical masses}

The dynamical masses are calculated using the following equation, utilising the FWHM of the CO line and assuming ordered circular rotation: 
\begin{equation}
      \hspace{0.8cm}  M_{\rm dyn} = 1.16\times10^5 \times 0.75 \times  FWHM_{\rm CO} \times sin(i)^2 \times r_{kpc}
\end{equation}

\noindent and assuming that the dark matter fraction is negligible and following the same method as was used for the WISSH survey \citep{Bischetti21}, also see \citep{Wang13, Venemans16}. We use i = 30 degrees as the average inclination assumed for these Type 1 quasars \citep{Mountrichas21}. We use r$_{\rm kpc}$ = 5\,kpc based on CO size measurements from \textsc{casa} when fitted using the \textsc{imfit} routine. Only three targets were spatially resolved and therefore have reliable size measurements. The sizes ranged between 0.53 and 0.65 arcsec, with an average of 0.6 arcsec, corresponding to 5\,kpc at these redshifts. We therefore assume 5\,kpc for all targets in the sample when calculating the dynamical masses. We discuss in Section~\ref{sec:discuss_gas_frac} the effect that this assumption has on the resulting gas fractions and gas masses and the differences obtained if using sizes of 2\,kpc \citep[as seen in ][]{Bischetti19, D'Amato20} in calculating the dynamical mass.

\begin{figure}
\centering
\includegraphics[width=\columnwidth]{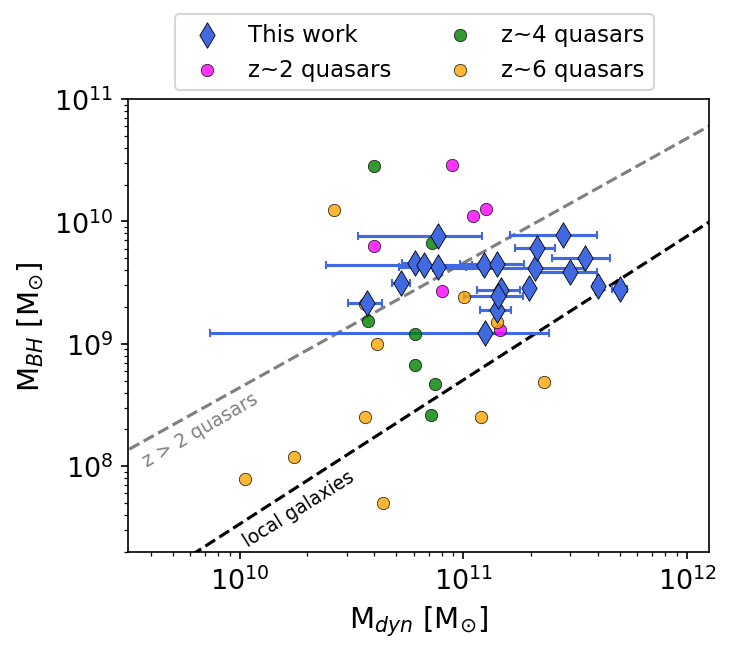}
\caption{M$_{\rm dyn}$ versus M$_{\rm BH}$ for all our targets with detections along with others from the literature. This includes quasars at z~$\sim$~2 from \citealt{Banerji15, Banerji17, Bongiorno14, Brusa18} and luminous z~$\sim$~4--6 QSOs from \citealt{Venemans16, Venemans17, Willott13, Willott15, Willott17, Kimball15, Trakhtenbrot17, Feruglio18, Mortlock11, DeRosa14, Kashikawa15} all of which are also collated in  \citealt{Pensabene20}. The dashed black line represents the local relation inferred from local galaxies \citep{Kormendy13}. The grey dashed line is the relation found by \citealt{Pensabene20} for a sample of quasars at $z > 2$. 
}
\label{fig:mdyn_Mbh}
\end{figure}

In Fig.~\ref{fig:mdyn_Mbh} we present the dynamical mass versus the black hole mass in comparison to other samples of quasars within the literature. We find that for this sample, the black holes are over-massive relative to the dynamical masses. The main reason may well be as a result of selection effects, since in this work we are selecting the most luminous quasars at these redshifts (see Fig.~\ref{fig:sample-select}). However, the results are consistent with the $z >$~2 relation identified in \cite{Pensabene21}. Likewise, over-massive black holes were identified in similarly luminous quasars in the WISSH survey \citep{Bischetti21}.
%\com{add stack spectra discussion, what did, test for outflows given figure 4. no broad}

\subsection{Gas masses and gas fractions}
\label{sec:calc_gas_frac}

To calculate the gas masses we first need to calculate the line brightness temperature, $L'_{CO} [{\rm K~ km~ s^{-1} pc^2}]$ using the following equation from \citealt{Solomon05}: 

\begin{equation}
\label{eq:linelum}
    %\hspace{0.8cm} \rm L'_{CO} [{\rm K~ km~ s^{-1} pc^2}] = \frac{3.25 \times 10^7}{\nu ^2_{\rm CO,rest}}\left(\frac{D^2_L}{1+z}\right)I
    \hspace{0.8cm} L'_{CO} = 3.25 \times~10^7 \ I_{\rm CO} \ \nu^{-2}_{\rm CO,obs} \ D^2_L \ (1+z)^{-3}
\end{equation}

in units of K km s$^{-1}$ pc$^2$, where $\nu_{\rm CO,rest}$ is the rest frequency of the CO line in GHz, $\rm D_L$ is the luminosity distance in Mpc, $z$ is the redshift and $\rm I$ is the velocity integrated line intensity measured in Jy \kms.

We then need to convert the CO(3-2) line luminosity to the CO(1-0) line luminosity and do so assuming a line ratio of $r_{31}$ = 0.97 \citep{Carilli13} which is applied in the following equation:

\begin{equation}
\label{eq:Lco_and_lineratio}
    \hspace{0.8cm}  L^\prime_{\rm CO(1-0)} =  \frac{L^\prime_{\rm CO(3-2)}}{r_{31}}
\end{equation}

For those sources with a different CO transitions we use the following line ratios: $r_{41}$ = 0.87 and $r_{71}$ = 0.2. These line ratio values were selected based on previous works studying similarly luminous quasars in the literature \citep{Bothwell13, Carilli13, Molyneux24}.\\

From the CO(1-0) line luminosity we calculate the gas mass $\rm M_{gas}$, using the following equation:

\begin{equation}
      \hspace{0.8cm}  M_{\rm gas} = \alpha_{\rm CO} \times L^\prime_{\rm CO(1-0)}
\end{equation}

where we use $\alpha_{\rm CO}$ = 0.8 M$_\odot$ / (K \kms pc$^2$) \citep{Downes98, Bolatto13}.\\

To calculate the stellar mass we use the following equation (adopted by \citealt{Nguyen20} and \citealt{Bischetti21} among others): 

\begin{equation}
\label{eq:mstar}
    \hspace{0.8cm}  M_\star = M_{\rm dyn} - M_{\rm BH} - M_{\rm gas}
\end{equation}

\noindent where the black hole mass is derived using log(L$_{5100}$) and FWHM$_{\rm Balmer}$ (measured from NIR spectral modelling in \citealp{Temple24}, and H$\beta$ black hole mass scaling relation from \citealp{Shen24}). Using the stellar mass and the gas mass we can calculate a gas fraction, f$_{\rm gas}$, using:

\begin{equation}
   \hspace{0.8cm}   f_{\rm gas} = M_{\rm gas} / M_\star
\end{equation}

Within our sample we find gas fractions ranging from 0.02 -- 0.32, and the values for each target can be found in Table~\ref{tab:CO_props}. We should caveat that since the stellar mass is calculated using the gas mass (as in equation~\ref{eq:mstar}), the gas fraction is dependent on the assumptions made to calculate the gas mass. However, using different line ratios and $\alpha_{\rm CO}$ to calculate the gas mass provided negative stellar masses in some cases and so the chosen values are considered reasonable assumptions.

\begin{figure}
\centering
\includegraphics[width=\columnwidth]{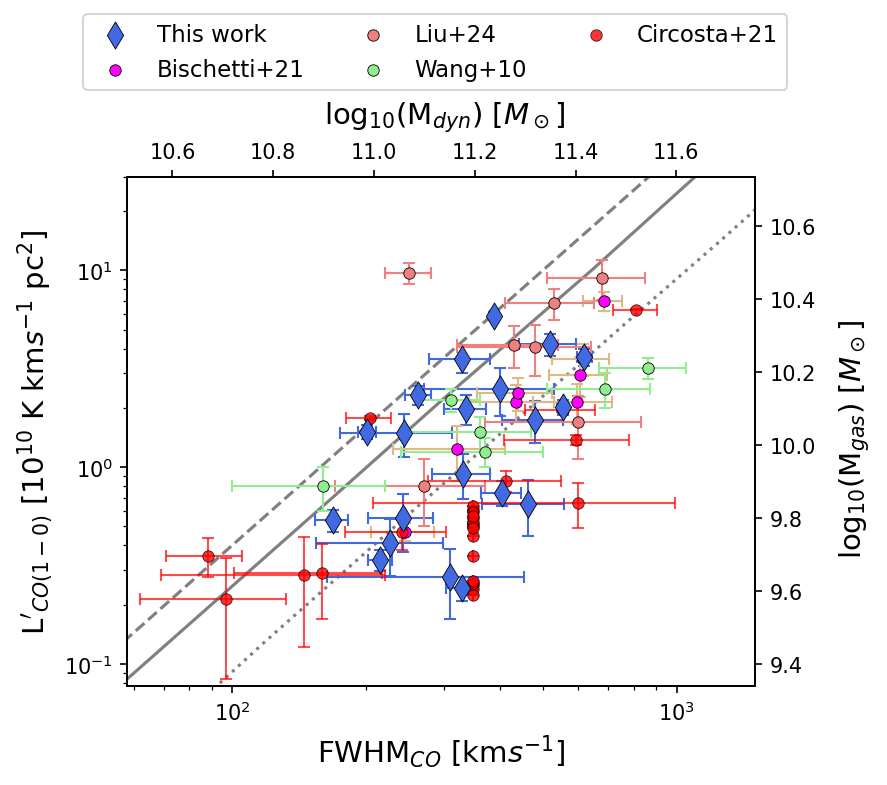}
\caption{Here we present the FWHM$_{\rm CO}$ versus the CO luminosity for all targets within our sample with detections. These properties trace the dynamical mass and the gas mass respectively. We plot other quasars and AGN taken from the literature \citep{Wang10, Circosta21, Bischetti21, Liu24}. The solid line corresponds to the approximate best-fitting quadratic relationship for submillimeter galaxies from \citealt{Bothwell13} (as also presented in \citealt{Liu24}). The dashed and dotted lines represent relations assuming spherical and disk models respectively \citealt{Liu24}.}
\label{fig:fwhm_co_vs_lco}
\end{figure}

In Fig.~\ref{fig:fwhm_co_vs_lco} we show the $FWHM_{\rm CO}$ versus the CO luminosity, which can be used as proxies for the dynamical mass and the gas mass respectively. Our findings are consistent with AGN taken from the literature and highlights low gas fractions within our sample (discussed in detail in Section~\ref{sec:discuss_gas_frac} as the dynamical masses are on average larger than the gas masses. The solid grey line represents the best fit for luminous sub-millimetre galaxies presented in \citealt{Bothwell13} and generally lies above our quasars. We also show relations assuming a disk model (dotted line in Fig.~\ref{fig:fwhm_co_vs_lco}) and a spherical model (dashed line in Fig.~\ref{fig:fwhm_co_vs_lco}). These models are formed from the following equation:

\begin{equation}
\hspace{2cm}    L ^\prime_{\rm CO(1-0)} = \frac{C(\Delta V / 2.355)^2R}{\alpha \cdot G} 
\end{equation}

\noindent where $\Delta$ V is the $FWHM$ of the CO line in \kms,R is the radius of the CO emission region in parsecs, $\alpha$ is the conversion factor from $L^\prime_{\rm CO(1-0)}$ to solar mass in K km s$^{-1}$ pc$^2$, G is the gravitational constant, and C is a constant related to the kinematics of the galaxy. As done in \citealt{Liu24} we use the following parameters from \citealt{Erb06}: C~=~2.1, R = 5 kpc, and $\alpha$~=~4.6 for a disk model; and C~=~5, R~=~2 kpc,and $\alpha$~=~1.0 for a spherical model. We find that our quasars are mostly consistent with virial relations assuming a disk model as opposed to a spherical model.

Given these low gas fractions we performed an analysis of a stacked spectrum of all CO(3-2) detections to test for the presence of underlying broad outflow components. However no broad component was identified, suggesting that outflows are not ubiquitous in this sample.

\subsection{Dust Continuum}

As with the analysis of the spectra (described in Section~\ref{sec:spectral_fitting}), the continuum images used here are the standard ALMA reductions. There are 18 continuum detected sources within the sample of 41 presented here (including archival data). Within the main sample we have a 38 per cent continuum detection rate (12 out of 32 with $>$ 3 sigma detection). From the archival CO(3-2) 2 out of 4 are detected. In the archival CO(4-3) data all 4 are detected in continuum (with much deeper observations). And finally in the archival CO(7-6) observation there was no continuum detection.

For those detected both in CO and in continuum, the emission was identified at consistent spatial locations. 4 targets also show continuum emission not associated with the quasar. For J2123$-$0050, emission is identified in the continuum which is co-spatial with CO emission from companions (see figure \ref{fig:cont_cutouts} and discussion on companions in Section~\ref{sec:companions_result}). %For the other sources that showed evidence for continuum emission not associated with the target quasar (J0052+0140, J0826+1434 and J1416+2649) we can't say whether or not these are from sources at the same redshift. and we analyse these using the \textsc{imfit} routine in \textsc{casa}. 

Five sources are found to have continuum emission at mJy levels, indicating they feature significant synchrotron emission. 
%Recent works have identified a tight correlation between the AGN and continuum emission \citep{Kawamuro22, Ricci23}. 
The median continuum flux for these sources is 2.65mJy, all of which were observed in the CO(3-2) transition. The remaining 13 continuum detected sources show continuum emission at a level expected as coming from the Rayleigh Jeans tail of the dust spectral energy distribution. These 13 sources were observed in both CO(3-2) and CO(4-3), and we find a median continuum flux for CO(3-2) and CO(4-3) observations of 128$\mu$Jy and 113$\mu$Jy respectively. For CO(3-2) observations, the continuum data was taken in the frequency range 97 -- 108 GHz. For CO(4-3) observations the continuum data taken in frequency range 141 -- 146 GHz. Finally, for the CO(7-6) observations the target had the continuum observed at a frequency of 237 GHz.

%\com{What do these levels of emission imply for the star formation rates of the quasar host galaxies?}

In comparison to the literature, we find a higher continuum detection rate than the SUPER survey (5 continuum detections out of 27) which study AGN at luminosities an order of magnitude lower than ours \citep{Circosta21}. In literature samples with similarly luminous quasars (but at higher redshifts) such as the WISSH survey \citep{Bischetti21}, a higher continuum detection rate is found with an 80 per cent detection rate, but similar continuum fluxes to our sample having a median of 0.25 mJy. Further, in a sample of Hot Dust-Obscured Galaxies (HotDOGs), again at similar luminosities and redshift, they find a higher detection rate, with 9 out of 13 detections \citep{Sun24}. They have a median continuum flux of 144$\mu$Jy for CO(3-2) detections, and so are slightly brighter in continuum in comparison to our sample, which is expected for luminous dust obscured quasars. For those with CO(4-3) detections in \citealt{Sun24}, the median continuum flux is 212$\mu$Jy, between 93 and 140 GHz, consistent with our CO(3-2) and CO(4-3) measurements.

We looked for correlations between the dust continuum properties of the quasars and their CO properties and found no significant dependencies. There were no differences in  C~\textsc{iv} blueshift, [O~{\sc iii}] $W_{80}$ and Eddington ratio between continuum detected and non-detected sources. Likewise, no difference in rest-frame UV or CO properties was found between quasars showing mJy level flux in the continuum and those showing $\mu$Jy level flux. All continuum flux densities are summarised in Table~\ref{tab:CO_props}.

%Number of continuum detected and also detected in CO = 11. \\
%Number of continuum detected but not detected in CO = 11.\\
%Number of non continuum and no CO = 9.\\
%Number of non continuum but CO detected = 10.\\

\subsection{Connecting the rest-frame UV quasar emission to the molecular ISM}
\label{sec:Correlations}

In this Section we investigate potential correlations between the global molecular gas properties (traced by CO) and the properties of the ionised gas in the BLR (traced by C~\textsc{iv} and He~\textsc{ii}) and in the NLR (traced by [O~{\sc iii}]), in this sample of high-luminosity, massive quasars. Previous works have identified a relation between outflow velocities in the quasar NLR and BLR, with [O~{\sc iii}] line widths correlating with C~\textsc{iv} blueshift \citep{Coatman19, Vietri18, Vietri20, Temple24} and so here we test whether a similar trend is observed in the molecular ISM. %One might expect that if [O~{\sc iii}] in the NLR correlates to C~\textsc{iv} then perhaps a similar trend would be observed in the molecular phase of the ISM.
We use only the main sample with CO(3-2) data, since they were observed to a uniform sensitivity and will therefore avoid any biases relating to the different observation setups. 
%Our goal is to study correlations between BLR/NLR outflow properties and the molecular gas for high-luminosity, massive quasars. 
%%%Manda%%%%
In our analysis we split the main sample into sub-samples based on the following properties: BH mass, Eddington ratio, C~\textsc{iv} blueshift, He~\textsc{ii} EW, [O~{\sc iii}] $W_{80}$ and BALs vs non-BALS. The specific values at which we split the sample are justified below.

\begin{figure*}
\centering
\includegraphics[width=\columnwidth]{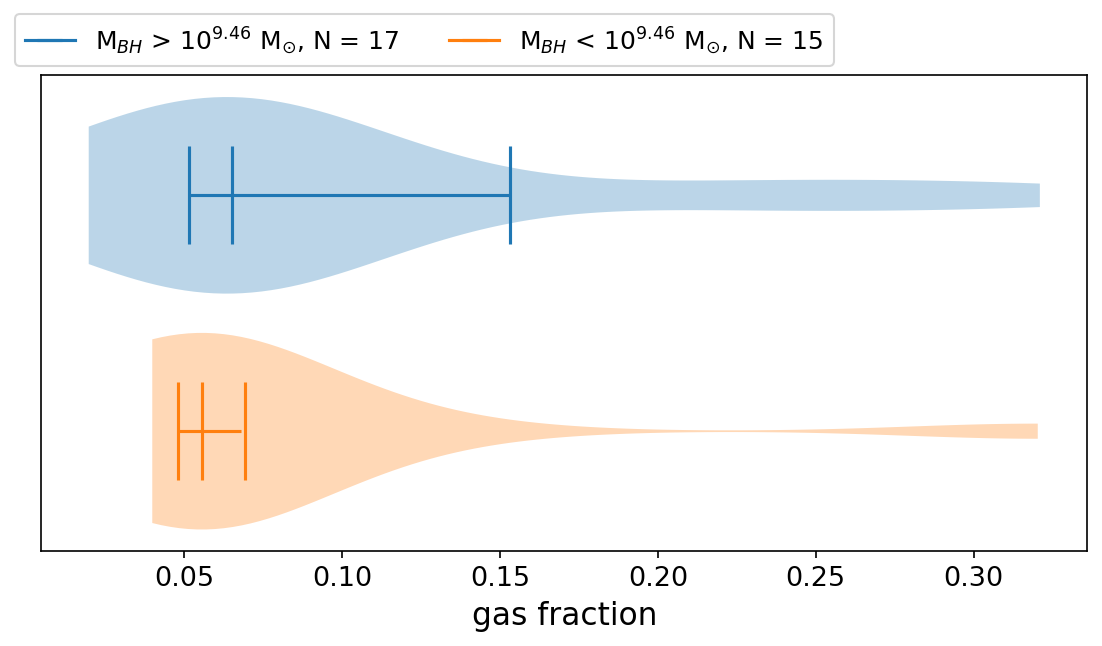}
\includegraphics[width=\columnwidth]{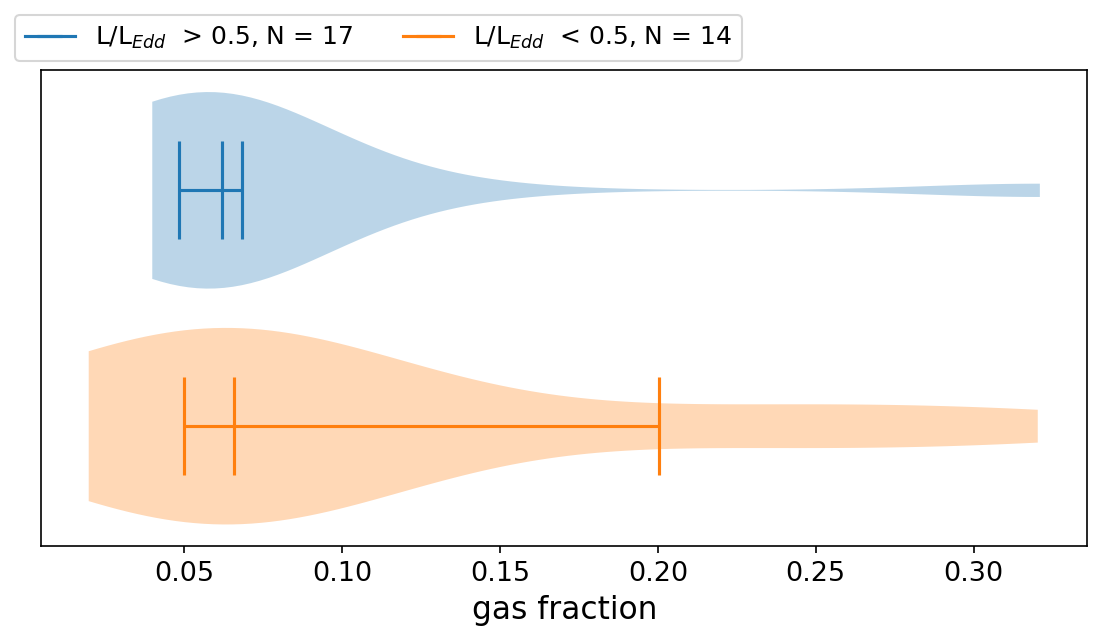}
\includegraphics[width=\columnwidth]{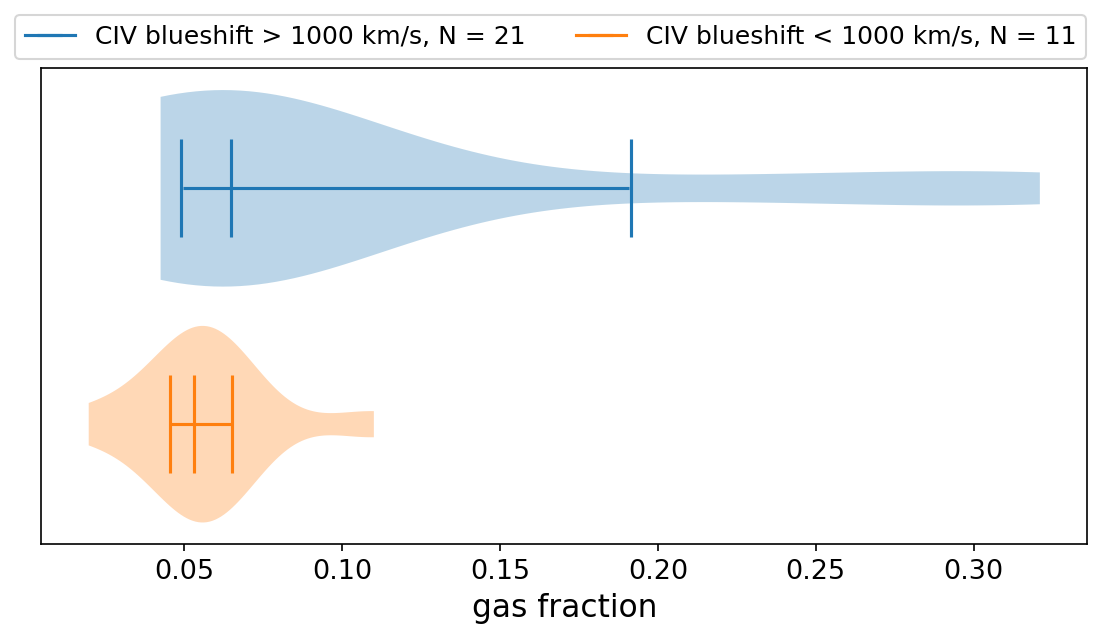}
\includegraphics[width=\columnwidth]{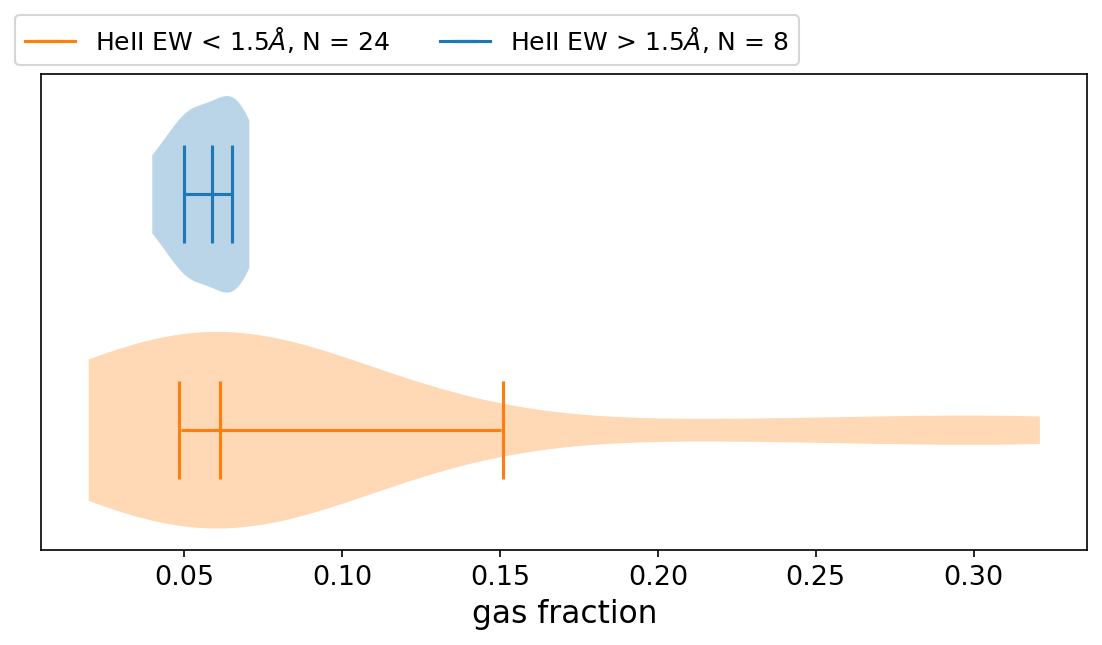}
\includegraphics[width=\columnwidth]{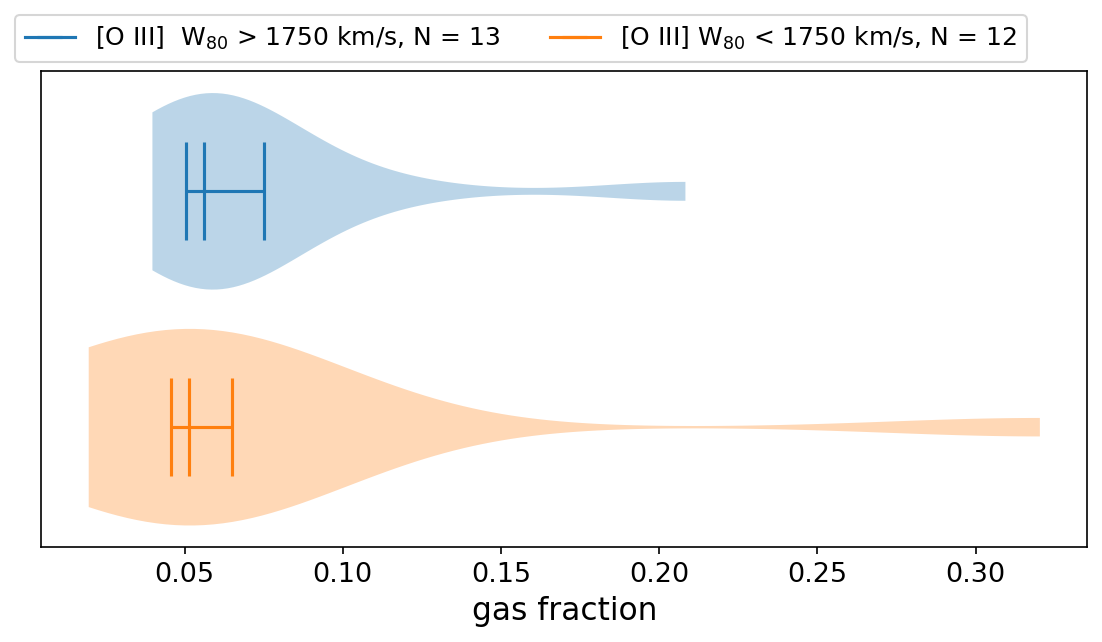}
\includegraphics[width=\columnwidth]{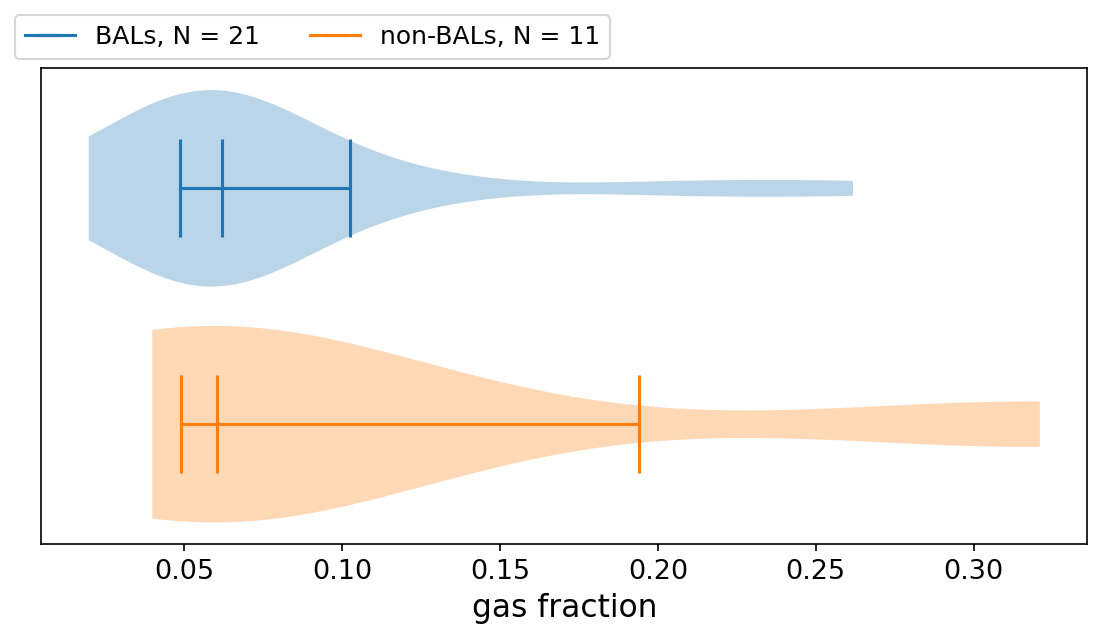}

\caption{Violin plots to show the distribution of gas fractions when splitting the sample by different UV/optical properties. The middle solid vertical lines represent the median of the distributions and the outer 2 vertical lines show 16th and 84th quartiles. The larger regions show the full distributions from minimum to maximum values. The wider the shaded region, the more data lie within that region. \textbf{Top row:} Splitting the sample by black hole mass of 10$^{9.46}$ M$_{\odot}$ (left) and by Eddington ratio = 0.5 (right). \textbf{Middle row:} Splitting the sample by C~\textsc{iv} blueshift = 1000 \kms (left) and He~\textsc{ii} EW  = 1.5 {\AA} (right) \textbf{Bottom row:} Split by [O~{\sc iii}] $W_{80}$ = 1750 \kms (left) and by BALs vs non-BALs (right). In all cases "N" is the number of sources within each distribution. Note that 3 sigma upper limits are also included here for those with non-detections.}
\label{fig:violin_plots_gas_frac}
\end{figure*}

Since the BLR/NLR outflow properties depend on both BH mass and Eddington ratio \citep{Temple23, Temple24}, we first split the sample roughly in half based on these two quantities, corresponding to $M_{BH}$ = 10$^{9.46}$ and $L/L_{Edd}$ = 0.5 (see top 2 panels in Figure~\ref{fig:violin_plots_gas_frac}). We confirm that there is no dependence of the gas fraction on BH mass and Eddington ratio. It should also be noted that the exact threshold values of BH mass and Eddington ratio chosen do not significantly change these distributions, which strengthens the evidence of no correlation. Having accounted for any potential dependence of the gas fraction on the BH mass and accretion rate, we can then test for differences in gas fraction between quasars with and without strong BLR winds corresponding to a C~\textsc{iv} blueshift threshold of 1000 \kms \citep{Temple23}. He~\textsc{ii} EW is anti-correlated with C~\textsc{iv}, so quasars with C~\textsc{iv} blueshifts $>$ 1000 \kms have weak He~\textsc{ii} lines with EW $<$ 1.5 {\AA} \citep{Rankine20, Temple23}. Splitting the sample at these values also evenly splits the population in the C~\textsc{iv} blueshift - equivalent width parameter space \citep{Richards11, Rivera22}. For [O~{\sc iii}], a $W_{80}$ of 1750 \kms represents the median for the sample in \citealt{Temple24}. 

After splitting the sample using these values, we analyse the distribution of gas fractions in each sub-sample to identify any differences. 
%CIV BS 1000km/s splits by strong winds in BLR versus no strong winds - refer to Temple+23. 
%HeII EW is anti-correlated with CIV BS  Splits the population in CIV space - Richards+11, Rivera+22).}
%https://ui.adsabs.harvard.edu/abs/2022ApJ...931..154R/abstract
%The value for each property where we split the data are physically motivated by the work carried out in \cite{Temple23, Temple24}, which shows differences in C~\textsc{iv} and He~\textsc{ii} properties across the M$_{\rm BH}$ and bolometric luminosity/Eddington ration parameter space.  
The distributions are presented by violin plots in Fig.~\ref{fig:violin_plots_gas_frac} and included within the distributions are 3 sigma upper limits for the non-detections (however removing these does not significantly alter the results). Based on the distributions there appears to be slight differences in the gas fractions based on the rest frame UV properties of the quasars. Quasars with higher C~\textsc{iv} blueshifts and softer ionising SEDs as traced by low EW He~\textsc{ii} emission seem to show a tendency to have higher gas fractions. The sub-samples of He~\textsc{ii} EW > 1.5 {\AA} and C~\textsc{iv} blueshift < 1000 km/s are made up of 100 per cent and 55 per cent non-detections respectively, which strengthens our tentative results. Given these are 3 sigma upper limits the gas fraction distributions in these sub-samples would be expected to be even lower, enhancing the differences. However, it should be noted that the differences currently are not statistically significant based on a two sided Kolmogorov–Smirnov test (KS test), with p-values of 0.06 (for C~\textsc{iv} blueshift and He~\textsc{ii} EW), where a value less than 0.05 is statistically significant. Nonetheless there are tentative signs of a dependence that would need investigating further with a wider sample to confirm. The remaining distributions, separating quasars based on other accretion and outflow properties (see Fig~\ref{fig:violin_plots_gas_frac}), show no statistical differences in their gas fractions when analysed with a KS test.

We must caution that with the number of sources analysed here it is hard to make robust conclusions, however we might be observing a fundamental link between the SED/accretion properties that is driving trends in the observed gas fractions. Quasars with stronger disk winds and softer ionising SEDs may preferentially be found in gas-rich host galaxies but more studies and statistical results are needed to confirm whether any correlation is real.

\subsection{No strong evidence for over-densities}
\label{sec:companions_result}

By analysing the presence of companions to the quasars in our sample, we can assess whether they are residing in overdense regions, which may suggest a larger supply of gas for fuelling AGN/quasar/galaxy growth. We identify companions that are clearly present within the same spectral window as the expected CO(3-2) frequency and that are offset spatially to the quasar. The field-of-view for the observations covers a diameter of 80 arcsec, corresponding to a physical distance of 670\,kpc at $z$~$\sim$~2, consistent with the searching radius with ALMiner (described in Section~\ref{sec:alma_obs}). Within the main sample of 32 quasars we identify three quasars with potential companions detected with SNR $>$ 3 sigma (J0104+1010, J1606+16735 and J2256+0105) and within the archival sample another three quasars show evidence of companions (see cutouts in Fig.~\ref{fig:all_spectra}).

Since identifying companions was not the main focus of this paper we did not perform a search for companions outside of the spectral window in which the CO was present. We therefore cannot rule out further companions with larger velocity offsets to the target quasar.

%\com{Another test performed to search for the presence of companions at the same redshift and in the local vicinity of our sample of quasars we analysed the spectra obtained at increasing radii. If the line intensity was seen to increase beyond normally expected (beyond 5kpc) we searched for brightness peaks within that radii. This lead to the finding of 2 further companions for J2256+0105 and J2108-0630. This analysis also proved as a test for the presence of extended diffuse gas surrounding the quasars. For 2 quasars: J0014+0912 and J1251+1143 we could identify no cause for the increasing line intensity beyond 5kpc and suggest low luminosity, diffuse emission to be the cause. }

%In J0104+1010, another component is identified at low signal-to-noise-ratio (S/N $\sim$ 2) but  found at the same velocities as the main component and are offset by $\sim$ 10 arcsec (corresponding to 80\,kpc). Given the S/N of these components, only one can be considered a marginal detection, with the second showing tentative signs of a detection that would need deeper observations to confirm.The observations for J0104+1010 reached sensitivities down to $\sim$~0.42mJy per 50~\kms bin width, which is slightly higher than the median for the sample which is 0.35mJy per 50~\kms bin width. For J1606+1735 we find no evidence for a detection at the location of the quasar, but find significant emission (S/N of 4.9) at a distance of $\sim$ 4 arcsec from the location of the quasar, corresponding to 32 kpc.

For the targets with archival data J0052+0104 has a companion which is 5 times brighter in CO(4-3) than itself, and is located $\sim$240\,kpc from the quasar. However, it should be noted that this detection was identified on the edge of the primary beam of the observations and so should be treated with some caution due to increased noise at the edge of the ALMA observations. J1416+2649 has at least two companions which are located at distances $\sim$ 30 and 75\,kpc from the quasar. J2123$-$0050 also shows two companions which are both comparably bright to the quasar and at distances of $\sim$60\ projected kpc from the quasar. The emission is both blue and redshifted either side of the quasar so this could also potentially be emission expelled from the host rather than companions. For more information on each of these individual targets see \cite{Li23}. %These observations go down to sensitivities of rms $\sim$ 0.2 mJy/beam compared to the rms of our CO(3-2) data of $\sim$ 0.3 mJy/beam.

%Some of the companions in the CO(4-3) data are also detected in the continuum data. Specifically, J2123-0050 shows evidence of both companions in the continuum, and the brighter companion to J1416+2649 seems also to be detected in the continuum (see figure \ref{fig:cont_cutouts}).

%\citep{Weiss07}

%\subsection{FWHM CO vs Lco}
%\label{sec:disc_fwhm_lco}
%\manda{input for this discussion would be appreciated if you can Manda}
%\com{The dotted lines represent the virial relations assuming spherical and disk models. High Lco for FWHM = spherical, low Lco for FWHM = disk} \citep{Bothwell13, Liu24}. 

%\com{Correlation between Lco and FWHM: traces of the mass of the gas reservoir (Lco) and the dynamics of the potential well in which it lies, dynamical mass and any inclination effects.}\\

\section{Discussion}
\label{sec:discussion}

Here we discuss our findings and interpretation of the results as well as putting them into context of the wider literature. In Section~\ref{sec:discuss_gas_frac} we discuss our gas fraction measurements and how they compare to literature samples across redshifts. In Section~\ref{sec:Gas_frac_obs_unobs} we discuss evidence for different gas fractions in obscured vs. unobscured quasars. Finally, in section~\ref{sec:disc_companions} we present a discussion about companions identified in the sample.

\subsection{Gas depletion in luminous Type 1 quasars at cosmic noon}
\label{sec:discuss_gas_frac}

\begin{figure*}
\centering
\includegraphics[width=0.9\textwidth]{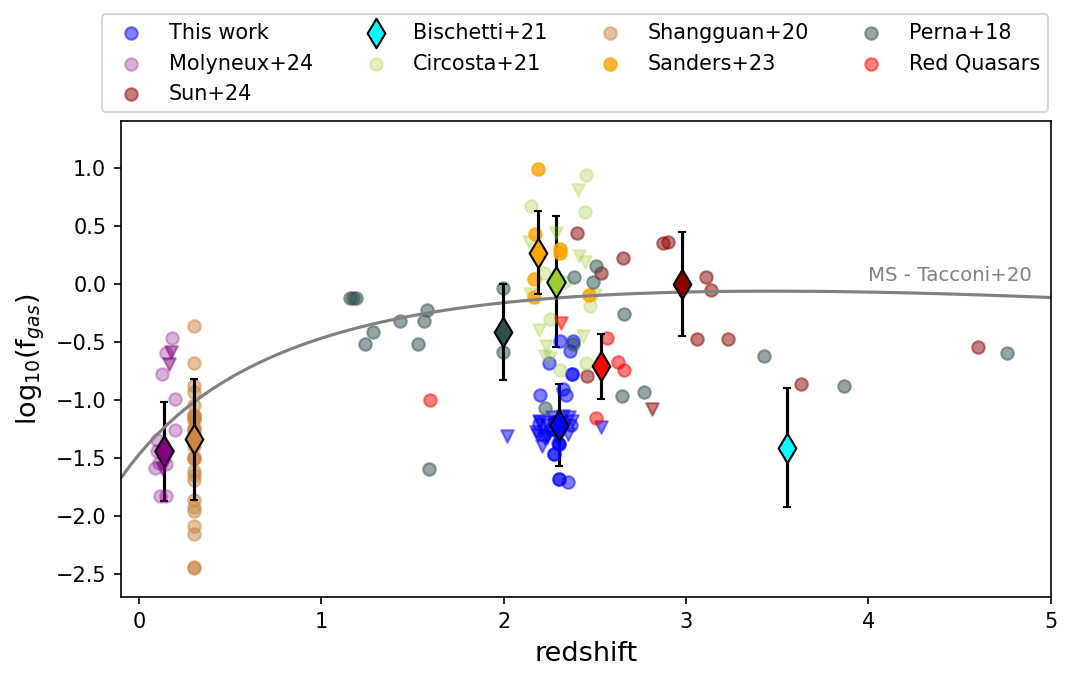}
\caption{Figure showing the gas fractions of our sample of quasars in relation to other samples of quasars, AGN and star-forming galaxies up to redshifts of 5. The grey line indicates the relation for main-sequence star-forming galaxies with stellar mass 10$^{11}$ M$_\odot$ described in \citealt{Tacconi20}. Upper limits are presented as downward triangles. For each sample presented the median redshift and gas fraction, as well as the standard deviation in gas fraction is presented as a diamond with the corresponding errorbars. Literature data taken from the following works: \citealt{Perna18, Banerji17, Banerji18, Shangguan20, Banerji21, Bischetti21, Circosta21, Sanders23, Molyneux24, Sun24}.
}
\label{fig:gas_frac_scaling}
\end{figure*}

As shown in Fig.~\ref{fig:gas_frac_scaling}, we present gas fractions for our sample alongside other samples of quasars, AGN and star-forming galaxies (SFGs) taken from the literature across the redshift range 0 -- 5. For this analysis we have assumed that each survey/presented work has chosen the most sensible values of $\alpha_{\rm CO}$ and line ratios based on their knowledge of their own samples. The different assumptions of line ratios and $\alpha_{\rm CO}$ for each literature sample can be found in Table~\ref{tab:lineratio_lit}. We identify similar gas fractions in our sample to the findings of the WISSH survey, which are similarly luminous quasars at $z > $2 \citep{Bischetti21}. However, we find lower gas fractions in our sources compared to lower luminosity quasars at the same redshift \citep[e.g.][]{Circosta21} and to SFGs also at the same redshift \citep{Sanders23}. We also find lower gas fractions when compared to a sample of HotDOGs at a slightly higher redshift \citep{Sun24}.

%In fact we find similar gas fractions to samples from $z~<$~0.5 \citep{Shangguan20, Molyneux24}, whereas other high-z SF galaxies are normally found with higher gas fractions than those identified in our work \citep[trend plotted from][]{Tacconi18}. 

Since we don't have direct stellar mass measurements and we instead derive them from the dynamical and gas masses (shown in equation ~\ref{eq:mstar}), we have analysed what values of $\alpha_{\rm CO}$ and line ratio would be required to increase our gas fractions to the level of other studies, namely \citealt{Perna18}, \citealt{Circosta21} and \citealt{Sun24}. If we keep $\alpha_{\rm CO}$ at 0.8 M$_\odot$ / (K \kms pc$^2$) we would need an r$_{31}$ of~$\sim$~0.1. Alternatively, fixing r$_{31}$ at 0.97 we would need an $\alpha_{\rm CO}$ of $\sim$ 11 M$_\odot$ / (K \kms pc$^2$). Both values seem unlikely for luminous quasars based on the literature. If we alter both parameters at the same time, we can match the gas fractions of our sample to the HotDOGs and SUPER sample using an $\alpha_{\rm CO}$ of~4.6 M$_\odot$ / (K \kms pc$^2$) and an r$_{31}$ of~0.4. Again, these values would seem unlikely given these are luminous quasars and that similarly luminous quasars in the literature are expected to have values of r$_{31}$~=~1 and $\alpha_{\rm CO}$~=~0.8 M$_\odot$ / (K \kms pc$^2$). We conclude that we are therefore indeed seeing depleted gas reservoirs compared to these literature samples. 

%\com{To match to SUPER and Sun gas fractions we would need a stellar mass 7\% the current calculated value. As our stellar mass is calculated from the dynamical mass and gas mass we can calculate what conversion of alpha co and r$_{31}$ would be needed to result in this value. } 

%In the literature, differences are observed in the gas fractions of AGN and non-AGN samples at redshift $\sim$ 2 \citep{Circosta21}. However, in low redshift samples evidence suggests that there is not such a large difference in the gas fractions. In particular, no differences were identified between AGN and non-AGN in \cite{Lamperti20}.

%There is also evidence in simulations that in the brightest quasars, the cold gas might be destroyed before it can be entrained in an outflow, leading to a tail-off in cold gas outflow rates at L$_{\rm bol}>$ 5x10$^{46}$ \citep[e.g.][]{Zubovas17, Ward24}. The sample presented in this work has L$_{\rm bol}>$ 10$^{46}$ so it would lie within this regime. This could therefore also be a contributing factor in the low gas fractions.

%Tacconi scaling relation across time. Differences between AGN and non-AGN. 

An important factor to consider in the gas masses, and therefore gas fractions, are the assumed sizes used to calculate the dynamical masses. As stated in Section~\ref{sec:calc_gas_frac} we assume 5\,kpc sizes since those which are resolved from the beam have these measured sizes. Molecular gas at 5\,kpc has been identified but as outflows and not part of the main bulk rotation  However, commonly quasars at these redshifts are found with sizes closer to 2\,kpc \citep{Bischetti19, D'Amato20}. If assuming 2\,kpc sizes, the median log$_{10}$(f$_{\rm gas}$) would be -0.76 $\pm$ 0.34 compared to -1.22 $\pm$ 0.27 when assuming r~=~5\,kpc. The overall result of depleted gas reservoirs in luminous quasars would therefore not change but the difference in gas fractions compared to the main sequence of star-forming galaxies would not be as large. 

Cosmological simulations such as SIMBA, EAGLE and IllustrisTNG suggest that AGN host galaxies at $z$~$\sim$~2 should match or exceed non-active galaxies in gas fraction \citep{Ward22}, which is the opposite of the observed trend. These simulations don't probe AGN and quasars with luminosities greater than $L_{\rm bol}$ = 10$^{45}$ erg s$^{-1}$ which are the luminosity of our quasars. For further discussion on the differences between simulations and observations see \citealt{Bertola24}.

\subsection{Gas fractions in obscured vs. unobscured quasars}
\label{sec:Gas_frac_obs_unobs}

\begin{figure}
\centering
\includegraphics[width=\columnwidth]{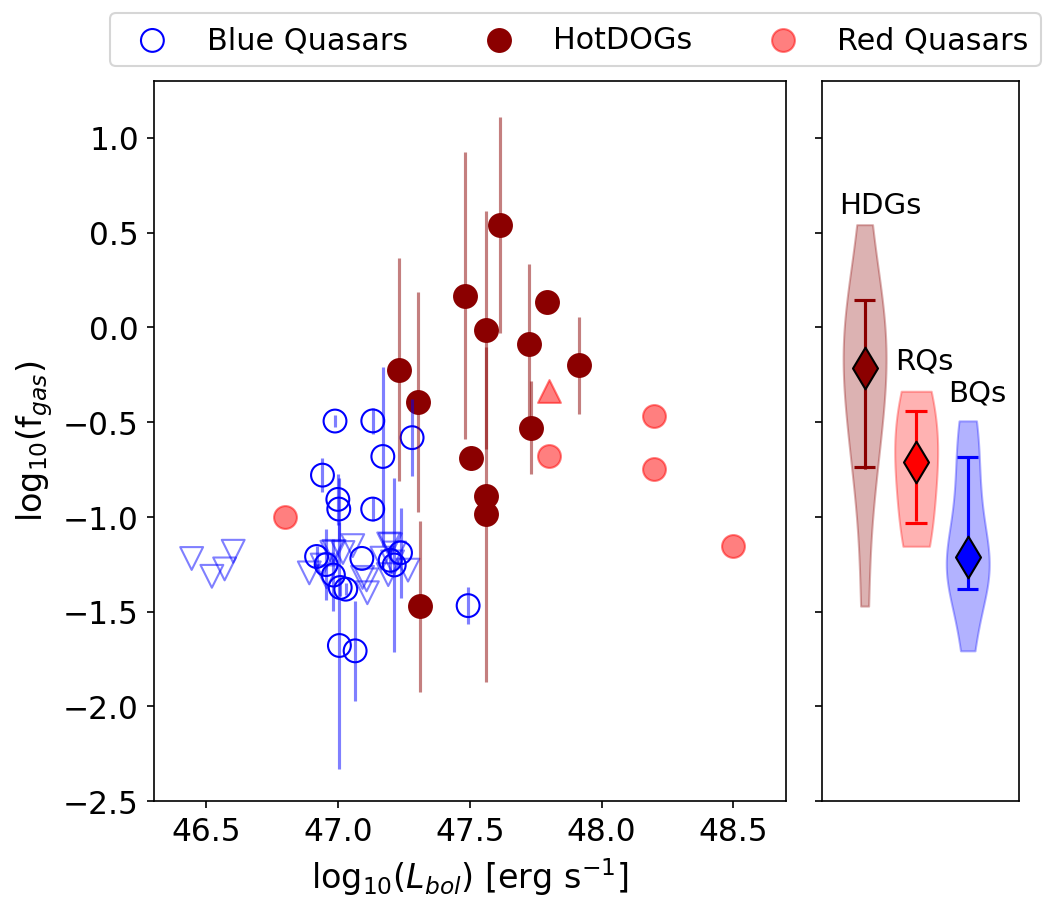}
\caption{Gas fraction vs bolometric luminosity. Here we focus on the AGN and quasars above z~=~2 and with L$_{\rm bol}$ $>$ 10$^{46.5}$ erg s$^{-1}$. Open circles and triangles are unobscured, filled circles and triangles are obscured/red quasars. For all samples presented here we calculate the gas fractions using the same $\alpha_{\rm CO}$ and line ratio for consistency. Blue quasars are data from this work. HotDOGs are data from \citealt{Sun24} and red quasars are compiled from \citealt{Feruglio14, Brusa15, Banerji17, Banerji18, Banerji21, Stepney24}.
}
\label{fig:gas_frac_lbol}
\end{figure}

In Fig.~\ref{fig:gas_frac_lbol} we compare the gas fractions in our sample of unobscured, blue quasars with luminous, obscured quasars above $z$~=~2 such as HotDOGs \citep{Sun24} and red quasars \citep{Banerji17,Banerji18,Banerji21}. We find lower gas fractions in our sample compared to both red quasars and HotDoGs, but with significant overlap between the samples. The HotDOGs have very high gas column densities \citep{Ricci17, Vito18, Assef20}, which could suggest that they are in an earlier evolutionary phase compared to our unobscured quasars. 

These results support the picture of dusty and red quasars, going through a 'blow-out' phase and evacuating the obscuring gas in the nuclear region, leaving more depleted gas reservoirs in the resulting blue quasars \citep[e.g.][]{Banerji12, Temple19, Lansbury20, CalistroRivera21}. In this picture, accretion exceeding the effective Eddington limit (due to the high fuel supply as a gas rich galaxy), is followed by an expulsion of material leading to a red quasar phase and finally an unobscured quasar phase. Recent observational results of red quasars may provide evidence of this blow-out phase, showing that molecules can survive in high velocity outflows even in quasars with $L_{\rm AGN}$ $>$ 10$^{48}$ erg s$^{-1}$ \citep{Stacey22b}. Such outflows are also predicted by simulations where radiation pressure on dust can launch outflows at galactic scales \citep{Costa18, Costa18b}. 

Overall, for this sample of luminous unobscured quasars, we conclude that we are observing host galaxies with lower gas fractions in comparison to similarly luminous obscured quasars at similar redshifts.

\begin{table*}
\begin{tabular}{|c|c|c|c|c|c|c|c|}
\hline
Sample                & Type       & redshift      & log(L$_{\rm bol}$)      & $\alpha_{\rm CO}$        & line ratio  & line sensitivity, $\Delta v$ = 10 \kms & beam size \\ 
                &        &       & (erg s$^{-1}$)      &     M$_\odot$ / (K \kms pc$^2$)    &   &  (mJy/beam)  & (arcsec) \\ \hline

This work             & Quasars    & 2.0 -- 2.6             & 46.4 -- 47.5            & 0.8                     & r$_{31}$ = 0.97  &   0.47 -- 0.77       & 0.5 - 3.0                \\ 
    &   &   &   &               & r$_{41}$ = 0.87  &       &               \\ 
        &   &   &   &               & r$_{71}$ = 0.20  &       &               \\ 
Sun+24                   & Hot DOGs    & 2.2 -- 4.6             & 47.1 -- 47.9    & 0.8                     & r$_{31}$ = 0.97      &     0.39 -- 1.40     & 0.3 -- 1.0                   \\ 
Bischetti+21                 & Quasars        & 2.4 -- 4.7             & 47.2 -- 48.0            & 0.8                     & r$_{41}$ = 0.87     &   0.38 -- 1.27 & 0.2 -- 5.0 \\ 
             &        &              &         &                    & r$_{51}$ = 0.69     &  \\ 
Red Quasars &  Quasars   & 1.5 -- 2.7 &   46.8 -- 48.5       & 0.8 &  r$_{31}$ = 1  &  & 0.8 -- 1.0          \\  
         Perna+18                 & Quasars  &      1.2 -- 4.8                &     43.9 -- 47.6   &  0.8 -- 3.6  & r$_{21}$ = 0.8 -- 1  & 0.17 -- 2.61 \\
                &  &                &         &         &  r$_{31}$ = 0.1 -- 1 &  \\
                &  &                 &        &   & r$_{41}$ = 0.41 -- 1 &  \\
Molyneux+24                & Quasars    & 0.1 -- 0.2 & 45.7 - 46.8              & 4                       & r$_{31}$ = 0.77 $^{+0.31}_{-0.20}$ $^{\cross[.3pt]}$  &  0.73 -- 2.76 & 4.3 -- 32.5              \\ 
Shangguan+20             & PG Quasars & 0.02 -- 0.1           & 44.7 -- 46.0            & 3.1                     & r$_{21}$ = 0.62                 &  3.57 -- 5.36 & 6.0 -- 8.0    \\ 
Circosta+21                 & AGN        & 2.1 -- 2.4             & 44.7 -- 46.9            & 3.6                     & r$_{31}$ = 0.5   &   0.46 -- 1.28   & 0.8 -- 1.7                  \\ 
Sanders+23               & SFGs       & 2.0 -- 2.5 &               & 0.66 -- 15.22 $^{\cross[.3pt]}$ &  r$_{31}$ = 0.55  &   & 0.7 -- 2.5          \\  
\hline
\end{tabular}
\caption{Table summarising the literature samples shown in Fig. \ref{fig:gas_frac_scaling} including type of object, redshift of the sample, bolometric luminosities, sensitivity of the observations as well as the $\alpha_{\rm CO}$ and line ratios used in those studies. Literature studies from the following works: \citealt{Perna18, Lamperti20, Shangguan20, Circosta21, Bischetti21, Sanders23, Molyneux24, Sun24}. Red Quasars are a compilation from \citealt{Feruglio14, Brusa15, Banerji17, Banerji18, Banerji21}. Line sensitivity quoted is for the upper CO transition, estimate taken from ALMA archive (or equivalent if not ALMA) to be consistent across all samples. Those with $^{\cross[.3pt]}$ are measured values as opposed to assumed values.}
\label{tab:lineratio_lit}
\end{table*}

%Molina                & AGN        & 0.3 -- 0.5 & 46.5 -- 47.2           & 3.1                     & r$_{21}$ = 0.62      &               \\ 
%Lamperti (BASS)       & AGN        & 0.0 -- 0.1 & 40 -- 47        & 4.35                    & r$_{31}$ = 0.52 $\pm$ 0.04 $^{\cross[.3pt]}$      &           \\ 
%Lamperti (xCOLD GASS) & SFGs       & 0.0 -- 0.1 & --              & 4.35 (1 for some) & r$_{31}$ = 0.53 $\pm$ 0.06 $^{\cross[.3pt]}$    &             \\ 
%Montoya Arroyave      & ULIRGs     & \textless~0.2 & \textgreater~46 & 1.7 $^{\cross[.3pt]}$                  & 0.76  $^{+0.22}_{-0.10}$      $^{\cross[.3pt]}$    &            \\ 
%Bothwell              & SMGs       & 1.2 -- 4.1    &        --         & 1                       &         0.52 $\pm$ 0.09 $^{\cross[.3pt]}$    &                      \\

\subsection{Companions to luminous quasars at z $\sim$ 2}
\label{sec:disc_companions}

Within the main sample of this work, with sensitivities down to $\sim$~0.5mJy per 33~\kms bin width, we find a few examples for CO-emitting companions to the quasars but no strong evidence that this is ubiquitous. As mentioned in Section~\ref{sec:companions_result}, three targets show strong evidence for companions - J0104+1010, J1606+1735 and J2256+0105. If there are companions for the other targets it would suggest that they are all less luminous than the detected quasars at these frequencies (rest frame frequency in the range $\sim$ 103 -- 108 GHz). 

%This is the case if there are companions present that are not luminous in CO(3-2) to be visible with the current observations, however there may simply be no companions present in the other targets.

The only other targets within the full sample of quasars presented here which have companions are those with archival CO(4-3) observations. These observations reached sensitivities $\sim$ 2 times better than for our sample (see Table~\ref{tab:obs_props}). These companions were first presented in \cite{Li23}. Out of the 10 quasars presented in \citet{Li23}, 7 had companions. It should be noted that the measured luminosities of the companions in these cases are mostly comparable to the observed quasar, and in only 1 of the 7 cases is the luminosity of the quasar much brighter than the companion(s). However, the CO(4-3) luminosities of the quasars in the sample of \cite{Li23} are significantly lower than the CO(3-2) luminosities in our sample, with a median of 4.4~$\times$~10$^9$~K~\kms~pc$^2$ compared to 17.5~$\times$ 10$^9$~K~\kms~pc$^2$ respectively (see Table~\ref{tab:CO_props} for the values for individual quasars).

In similarly luminous quasars to ours from the WISSH survey at $z > 2$, 80 per cent of the quasars were identified as residing in high density environments, with the sample showing line emitting companions at distances between 6 and 130\,kpc from the quasar \citep[][]{Bischetti21}. Within the literature there are many other examples of quasar systems with over-densities detected \citep[e.g.][]{Banerji17, Banerji18, Diaz-Santos18, Fogasy20, Lambert24, Zewdie24}. These also include samples of quasars with intense ongoing star formation (SFR $>$ 1000 M$_{\odot}$ per year). Spectroscopically identified companions to luminous quasars are found up to redshifts of 6 \citep[e.g.][]{Trakhtenbrot17, Decarli17}. Companions have also been found to be as bright as the quasar host galaxies themselves indicating that these galaxies and quasars are residing in dense, gas rich environments \citep{Neeleman19}.

We can also look to simulations to provide context on the predictions that luminous quasars should reside in over-densities. For example, GALFORM simulations predict that 33 per cent of simulated quasars at $z$~=~2.8 would have companion galaxies at distances $<$ 350\,kpc \citep{Fogasy17, Fogasy20}. Of these simulated quasars, 2.4 per cent would be expected to have a bright companions with SFR $>$ 100 M$_{\odot}$ per year. These simulations were analysed with a time slice of $\sim$ 200 Myr, corresponding to $\Delta$$z$ $\sim$ 0.2. 

The findings of our work suggests one of two things: either there are only a few cases for companions present in the main sample of quasars and therefore we find no strong evidence for over-densities. Alternatively, there are companions present but we would need to go much deeper with observations to identify them, as was done for the CO(4-3) work \citep{Li23}.

%The area covered by our ALMA observations covers a diameter of $\sim$330 kpc (at the median redshift of z = 2.3), a very similar radius as used in analysis of the GALFORM simulations. Since we may only be able to see the brightest companions due to sensitivity limits we might expect similar results to the lower percentage from GALFORM. With only one potential sign of companion(s) out of the main sample of 32 quasars, this would give a $\sim$3 per cent companion rate, which would indeed be a close match to the simulations. 

\section{Conclusions}
\label{sec:conclsions}

In this work we have presented ALMA CO observations of a sample of 41 quasars at $z~\sim~2$, with 32 sources observed in CO(3-2) and 9 archival sources with either CO(3-2), CO(4-3) and CO(7-6) observations. Most of the sources are spatially unresolved with only three being marginally resolved. The observations therefore trace the global molecular ISM properties in these quasars. We compare the gas fractions of our sample to literature samples of quasars/AGN and non-AGN across the redshift range 0 -- 5, as well as specifically to red quasars and HotDOGs. We compare the observed molecular gas properties to available multi-wavelength data tracing ionised gas on both sub-parsec and kpc scales (C~\textsc{iv} and [O~{\sc iii}]). The main findings from this work are as follows: \\

\begin{itemize}

\item{We obtain a 47 per cent detection rate in the main sample of sources with CO(3-2) data. From the nine sources included from the ALMA archive, we find 4 non-detections in CO(3-2), 4 detections in CO(4-3) and a single detection in CO(7-6). We identify a range of CO properties and find a median gas mass of 8.0~$\pm$~1.5~$\times$~10$^9$~M$_{\odot}$ and a range of $FWHM_{\rm CO}$ from 200 $\sim$ 600 \kms.}

\item{Our observations of molecular gas in a large, statistically significant sample at $z\sim2$ reveals evidence for depleted gas reservoirs in luminous, unobscured $z$~=~2 quasars with a median gas fraction of 0.06~$\pm$~0.09. This is compared to similarly luminous red quasars and HotDOGs in the literature with gas fractions of 0.28~$\pm$~0.13 and 0.63~$\pm$~0.90 respectively. This indicates a trend of increasing gas fraction with obscuration and may support the idea of an evolutionary phase in AGN, where quasar host galaxies move from more gas rich and obscured to gas poor and unobscured.}

\item{We identify tentative hints at a correlation between the gas fractions and He~\textsc{ii} EW and C~\textsc{iv} blueshifts. Quasars with C~\textsc{iv} blueshifts $>$ 1000 \kms have gas fractions in the range 0.04 -- 0.32  compared to those with C~\textsc{iv} blueshifts $<$ 1000 \kms with lower gas fractions in the range 0.02 -- 0.11. Similarly quasars with He~\textsc{ii} EW $>$ 1.5 \AA\ have gas fractions in the range 0.04 -- 0.07 whereas those with He~\textsc{ii} EW $<$ 1.5 \AA\ have gas fractions in the range 0.02 -- 0.32. These correlations are currently not statistically significant and larger samples are needed to test this further. High C~\textsc{iv} blueshift quasars with softer ionising SEDs (and therefore lower He~\textsc{ii} EW) are expected to be driving stronger disk winds in the quasar broad line region. Our results may therefore point to a link between strong BLR outflows and enhanced gas fractions in luminous quasars at cosmic noon.}

\item{Three targets out of the main sample of 32 show signs of a companion galaxy detected in CO. ALMA archival data with deeper observations show more significant evidence for companions, with 3 out of 4 observed in CO(4-3) showing evidence for companions. This suggests that lower luminosity companions might be present across the main sample, however we would need deeper observations to test this further. Overall we find no strong evidence for over-densities in our sample of luminous quasars at $z\sim2$.}

\end{itemize}

%Overall, we find low gas fractions in this sample of luminous, unobscured quasars at $z\sim$2, and find tentative correlations between these gas fractions (and CO properties) to the properties of the ionised gas on sub-parsec and kpc scales. Future observations with larger samples and corresponding rest frame UV data as well as deeper observations with ALMA of the molecular gas are required to increase the statistics for these objects and to identify the cause of the measured low gas fractions. 

%%%%%%%%%%%%%%%%%%%%%%%%%%%%%%%%%%%%%%%%%%%%%%%%%%

\section*{Acknowledgments}

SJM acknowledges funding from The Royal Society through a Research Fellows Enhancement Award Grant (RF/ERE/221053) awarded to MB. MB acknowledges funding from The Royal Society via a University Research Fellowship. MJT was supported by UKRI STFC (ST/X001075/1).  CR acknowledges support from Fondecyt Regular grant 1230345, ANID BASAL project FB210003 and the China-Chile joint research fund. ALR acknowledges support from a UKRI Future Leaders Fellowship (grant code: MR/T020989/1). GCJ acknowledges support by the Science and Technology Facilities Council (STFC), by the ERC through Advanced Grant 695671 ``QUENCH'', and by the UKRI Frontier Research grant ``RISEandFALL.'' RJA was supported by FONDECYT grant number 1231718 and by the ANID BASAL project FB210003. MA acknowledges support from ANID Basal Project FB210003 and and ANID MILENIO NCN2024\_112. For the purpose of open access, the author has applied a Creative Commons Attribution (CC BY) licence to any Author Accepted Manuscript version arising from this submission.

\section*{Data Availability}

This paper makes use of the following ALMA data: ADS/JAO.ALMA$\textbackslash\#$2021.1.00393.S, ADS/JAO.ALMA$\textbackslash\#$2016.1.00798.S, ADS/JAO.ALMA$\textbackslash\#$2013.1.01262.S, ADS/JAO.ALMA$\textbackslash\#$2017.1.01676.S, ADS/JAO.ALMA$\textbackslash\#$2019.1.01251.S, ADS/JAO.ALMA$\textbackslash\#$2018.1.00583.S. ALMA is a partnership of ESO (representing its member states), NSF (USA) and NINS (Japan), together with NRC (Canada), NSTC and ASIAA (Taiwan), and KASI (Republic of Korea), in cooperation with the Republic of Chile. The Joint ALMA Observatory is operated by ESO, AUI/NRAO and NAOJ."

The data used in this work are available from the ESO Science
Archive Facility (https://archive.eso.org/). The reduced data underlying this paper will be shared on reasonable request to the corresponding author.

\onecolumn
\setcounter{table}{1}
\begin{table*}
\centering
\begin{longtable}{|ccccccccc|}
\hline
\textbf{Source}  & \multicolumn{1}{c}{\textbf{CO Line}} & \multicolumn{1}{c}{\textbf{z$_{\rm CO}$}} & \multicolumn{1}{c}{\textbf{Line intensity}} & \multicolumn{1}{c}{\textbf{FWHM$_{\rm CO}$}} & \multicolumn{1}{c}{\textbf{$\rm L'_{CO}$}} & \multicolumn{1}{c}{\textbf{log(M$_{\rm gas}$})} & \multicolumn{1}{c}{\textbf{gas fraction}} & \multicolumn{1}{c|}{\textbf{S$_{\rm cont}$}} \\ 
     &    &      &  [mJy km~$\rm s^{-1}$]   &     [km~$\rm s^{-1}$]          &   [1x10$^9$ $\times$ K \kms pc${^2}$]           &   [M$\odot$)]              & & [mJy]             \\
     (1)     &  (2)  &  (3)    &  (4)   &     (5)    &   (6)    &   (7)       &     (8)      & (9)  \\ \hline
J0014+0912      & CO(3-2)    & 2.3478 $\pm$ 0.0003      &  1460 $\pm$ 186   &     518 $\pm$ 76          &      42.31 $\pm$  5.38      &     10.54$\pm$   0.05        &      0.11 $\pm$ 0.04   & $<$ 0.03 \\ 
J0019+1555      & CO(3-2)    &  --    &   $<$ 367       &       --        &     $<$  10.49       &       $<$  9.92         &    $<$   0.07              & 0.50 $\pm$ 0.03 \\ 
J0052+0140      & CO(4-3)    &  2.3104 $\pm$ 0.0001    &   205 $\pm$ 25   &      215 $\pm$ 13         &      3.25 $\pm$ 0.39       &   9.49 $\pm$  0.05            &    0.06 $\pm$ 0.01   &  0.11 $\pm$ 0.02   \\ 
J0104+1010      & CO(3-2)    & 2.3634 $\pm$ 0.0002   &   291 $\pm$ 66   &      262 $\pm$ 18        &      8.53 $\pm$ 1.93      &       9.83 $\pm$  0.05        &     0.06 $\pm$ 0.06        &  0.27  $\pm$ 0.03   \\ 
J0105+1942      & CO(3-2)    & 2.3232  $\pm$ 0.0002  &   667 $\pm$ 113   &     336 $\pm$ 37          &     18.99 $\pm$ 3.22       &      10.20 $\pm$ 0.07          &    0.12 $\pm$ 0.04           & $<$ 0.05       \\ 
J0106+1010      & CO(3-2)    &  --    &   $<$ 279      &       --        &    $<$  8.45         &      $<$  9.83          &     $<$    0.05        &  $<$ 0.02    \\ 
J0106$-$0315      & CO(3-2)    &  2.2412 $\pm$ 0.0003  &   533 $\pm$ 131      &   243 $\pm$ 69            &   14.25 $\pm$ 3.50          &    10.08 $\pm$ 0.10            &        0.21 $\pm$ 0.16     &  83.7 $\pm$ 1.60   \\
J0140$-$0138      & CO(3-2)    &  --    &   $<$ 358       &       --        &      $<$ 9.99       &       $<$ 9.90          &     $<$    0.06         &  $<$ 0.03   \\ 
J0142+0257      & CO(3-2)    &  --    &   $<$ 295       &       --        &      $<$ 8.71       &      $<$  9.84         &         $<$    0.52        & 1.64 $\pm$ 0.03 \\ 
J0229$-$0402      & CO(3-2)$^a$    &  --    &   $<$ 405      &        --       &     $<$  10.88      &     $<$  9.94           &     $<$    0.07   &  1.97  $\pm$ 0.04       \\ 
J0351$-$0613      & CO(3-2)    &  --   &   $<$ 283       &       --        &      $<$  7.80      &      $<$   9.80         &    $<$     0.05          &  0.59 $\pm$ 0.02 \\ 
J0758+1357      & CO(3-2)    &  --    &   $<$ 246      &          --     &      $<$  6.69      &      $<$   9.73         &     $<$     0.04        &    2.65 $\pm$ 0.04\\ 
J0810+1209      & CO(3-2)    &  --    &   $<$ 406      &      --         &      $<$  11.59      &     $<$   9.97          &     $<$     0.07     &   $<$ 0.03    \\ 
J0811+1720      & CO(3-2)    &  --   &   $<$ 367     &        --       &       $<$  10.93     &       $<$   9.94        &      $<$     0.07   & $<$ 0.05       \\ 
J0815+1540      & CO(3-2)    &  --    &   $<$ 288    &          --     &       $<$  8.01    &        $<$  9.81       &       $<$     0.05     &  $<$ 0.04   \\ 
J0826+1434      & CO(3-2)    &  --   &   $<$ 356    &           --    &        $<$  10.50    &      $<$  9.92          &        $<$    0.06    &   2.71  $\pm$ 0.04  \\ 
J0826+1635      & CO(3-2)    &  --    &   $<$ 309    &        --       &       $<$  8.29     &       $<$  9.82         &         $<$   0.05    &   0.10 $\pm$ 0.02   \\ 
J0827+0618      & CO(3-2)    &  --   &   $<$ 392    &         --      &        $<$  10.59    &      $<$  9.93          &        $<$    0.06     &   $<$ 0.13   \\ 
J0832+1823      & CO(3-2)    &  2.2785 $\pm$ 0.0003  &   190 $\pm$ 62    &     242 $\pm$ 40          &  5.23 $\pm$ 1.71          &      9.64 $\pm$ 0.15          &     0.06 $\pm$ 0.03      &  $<$ 0.03          \\ 
J0837+0521      & CO(3-2)    &   --   &   $<$ 377    &      --         &      $<$  11.52      &    $<$  9.96            &         0.15          & $<$ 0.04  \\ 
J1000+0206      & CO(3-2)$^a$    &   --   &   $<$ 344    &       --        &      $<$  9.94      &    $<$  9.90            &       $<$ 0.07        & $<$ 0.05    \\ 
J1006+0119      & CO(7-6)    &  2.3096 $\pm$ 0.0001  &   451 $\pm$ 64    &    329 $\pm$ 26           &   2.33 $\pm$ 0.33         &   9.74 $\pm$ 0.06         &     0.06 $\pm$ 0.01       &  $<$ 0.07      \\ 
J1113+1022      & CO(3-2)    &  2.2699 $\pm$ 0.0003  &   145 $\pm$ 46    &    226 $\pm$ 72           &   3.96 $\pm$ 1.26         &     9.52  $\pm$ 0.14          &     0.06 $\pm$ 0.04           & $<$0.02      \\ 
J1213+0807      & CO(3-2)    &   --   &   $<$ 347      &       --        &    $<$  10.75        &       $<$ 9.93          &     $<$    0.07          &  $<$ 0.03 \\ 
J1251+1143      & CO(3-2)    &  2.2003 $\pm$ 0.0004  &   965 $\pm$ 262      &   400 $\pm$ 130      &  24.99 $\pm$ 6.78   &  10.31  $\pm$ 0.11     &  0.11 $\pm$ 0.08         &  0.09 $\pm$ 0.02  \\ 
J1416+2649      & CO(4-3)    &  2.2997 $\pm$ 0.0003  &   167 $\pm$ 65      &    309 $\pm$ 145           &  2.63  $\pm$ 1.02         &     9.40 $\pm$ 0.18          &     0.02 $\pm$ 0.02     &  0.41 $\pm$ 0.02   \\ 
J1420+1603      & CO(3-2)$^b$    &   --   &   $<$ 269      &       --        &     $<$ 8.18        &     $<$   9.82         &    $<$      0.05       & $<$ 0.02     \\
J1532+1739      & CO(3-2)    &  2.2006 $\pm$ 0.0006  &   214 $\pm$ 68      &    462 $\pm$ 97           &   5.54 $\pm$ 1.76         &       9.72 $\pm$  0.14        &       0.02 $\pm$ 0.01         &   $<$ 0.01    \\ 
J1606+1735      & CO(3-2)    &  -- &   531 $\pm$ 189      &    540 $\pm$ 120           &   15.85 $\pm$ 5.64         &      10.10 $\pm$ 0.16          &    $<$ 0.06        &  $<$ 0.01         \\ 
J1625+2646      & CO(3-2)$^c$    &   --   &   $<$ 291    &        --       &       $<$ 8.58    &       $<$  9.84         &      $<$  0.06     &  0.17 $\pm$ 0.03     \\ 
J2059$-$0643      & CO(3-2)    &  2.3327 $\pm$ 0.0001  &   511 $\pm$ 41    &  201 $\pm$ 9             &  14.65  $\pm$ 1.18         &       10.08  $\pm$ 0.04        &        0.32 $\pm$ 0.05      &  $<$ 0.05      \\ 
J2108$-$0630      & CO(3-2)    &  2.3344 $\pm$ 0.0001  &   1154 $\pm$ 168      &   329 $\pm$ 52            &  33.13 $\pm$ 4.82          &    10.45  $\pm$ 0.06           &        0.26 $\pm$ 0.05        & $<$ 0.05      \\ 
J2121+0052      & CO(4-3)    &  2.3736 $\pm$ 0.0001  &   309 $\pm$ 39    &    168 $\pm$ 14           &    5.14  $\pm$ 0.65       &    9.69  $\pm$ 0.06           &      0.17 $\pm$ 0.03       & 0.06 $\pm$ 0.01   \\ 
J2123$-$0050      & CO(4-3)    &  2.2813 $\pm$  0.0002 &   459 $\pm$ 66   &     404 $\pm$ 41          &     7.12  $\pm$ 1.02      &      9.83  $\pm$ 0.06         &       0.03 $\pm$ 0.01       &    0.12 $\pm$ 0.02  \\ 
J2239$-$0047      & CO(3-2)    &  2.2234 $\pm$ 0.00023  &   635 $\pm$ 155   &     480 $\pm$ 75          &    16.75 $\pm$ 4.09        &      10.15 $\pm$ 0.11          &     0.05 $\pm$ 0.02        &  $<$ 0.09    \\ 
J2256+0105      & CO(3-2)    &  2.2716 $\pm$ 0.0003  &   325 $\pm$ 83         &  330 $\pm$ 49            &   8.90 $\pm$ 2.27              &  9.87 $\pm$ 0.11   &   0.06 $\pm$ 0.02     &  $<$ 0.04   \\ 
J2256+0923      & CO(3-2)    &  2.2981 $\pm$ 0.00033  &   693 $\pm$ 63   &   554 $\pm$ 15            &     19.36 $\pm$ 1.76       &       10.20  $\pm$ 0.04        &       0.04 $\pm$ 0.01      &  $<$ 0.05         \\
J2300+0031      & CO(3-2)    &   --   &   $<$  327   &     --          &     $<$ 8.66        &      $<$ 9.84           &     $<$    0.05        &  0.08 $\pm$ 0.02 \\ 
J2314+1824      & CO(3-2)    &   --   &   $<$  277    &    --           &      $<$ 8.04        &     $<$ 9.81            &     $<$    0.05       &  $<$ 0.06     \\ 
J2348+1933      & CO(3-2)    &  2.1932 $\pm$ 0.0002  &    1345 $\pm$ 129   &    619 $\pm$ 24           &   34.63 $\pm$ 3.32         &      10.46  $\pm$ 0.04         &       0.06 $\pm$ 0.01     &  0.18  $\pm$ 0.03   \\ 
J2352$-$0120      & CO(3-2)    &  2.3826 $\pm$  0.0001 &    1888 $\pm$ 83   &    388 $\pm$ 7           &    56.15 $\pm$ 2.47         &     10.67   $\pm$ 0.02         &    0.32 $\pm$ 0.02        &   0.12 $\pm$ 0.02     \\ \hline
Companions      &    &  &      &       &     &       &   &    \\
J0104+1010  & CO(3-2)      & 2.3632 $\pm$ 0.0004 & 337 $\pm$ 93     &   350 $\pm$ 112  &   9.88 $\pm$ 2.72  &    --   &   -- &   -- \\ 
J1606+1735      &  CO(3-2)  & 2.3327 $\pm$ 0.0003 &  489 $\pm$ 106    &  517 $\pm$ 130     &   14.02 $\pm$ 3.04  &   --   &  -- & --   \\ 
%J0104+1010 $\#$ 2 &  CO(3-2)      & 2.3635 $\pm$ 0.0007 &   134 $\pm$ 64  &  341 $\pm$ 189     &   3.93 $\pm$ 1.88  &  --   & --  &  --  \\ 
J2108$-$0630 &  CO(3-2)  & 2.3793 $\pm$ 0.0004 & 236 $\pm$ 71   &   310 $\pm$ 108  &  7.02 $\pm$ 2.12  &     -- & --  &  --  \\
J2256+0105 &  CO(3-2)  & 2.2735 $\pm$ 0.0004 &  353 $\pm$ 70    &   486 $\pm$ 111    &   9.70 $\pm$ 1.90  &    --  & --  &  --  \\
J0052+0140     &  CO(4-3)      & 2.3097 $\pm$ 0.0002     &  756 $\pm$ 133     & 332 $\pm$ 73    &  11.98 $\pm$ 2.11     & -- &  -- &  --\\ 
J1416+2649 $\#$ 1 & CO(4-3)        &  2.2932 $\pm$ 0.0005 &  359 $\pm$ 62    &   586 $\pm$ 118    &   5.62 $\pm$ 0.97  &  --  &  -- &  --  \\ 
J1416+2649 $\#$ 2 &  CO(4-3)        &  2.2898 $\pm$ 0.0001 &  83 $\pm$ 18    &  121 $\pm$ 31     &   1.30 $\pm$ 0.28  & -- & --  &  --  \\
J2123$-$0050 $\#$ 1 & CO(4-3)        & 2.2844 $\pm$ 0.0003 &  370 $\pm$ 50    &   495 $\pm$  77    &  5.75 $\pm$ 0.78   & --  &  -- &  --  \\ 
J2123$-$0050 $\#$ 2 & CO(4-3)  & 2.2778 $\pm$ 0.0003   &    367 $\pm$ 54  &   477 $\pm$ 81    & 5.68 $\pm$ 0.84    & --   &  -- &  --  \\ 
\hline
\end{longtable}
\caption{CO line properties: (1) SDSS source name. (2) CO transition used. $^a$ indicates CO(3-2) from SUPER survey \citealt{Circosta21} and $^b$, $^c$ from 2013.1.01262.S and 2017.1.01676.S  respectively, not the main CO(3-2) sample here. (3) CO redshift determined from the V$_{50}$ and corresponding uncertainty. (4) line intensity. (5) FWHM of CO line and corresponding uncertainty. (6) CO line luminosity (brightness temperature) for the specific CO transition observed (7) Gas mass, derived from the CO line luminosity and assuming line ratios of $r_{31}$ = 0.97, $r_{41}$ = 0.87 and $r_{71}$ = 0.2, and $\alpha_{\rm CO}$ = 0.8 M$_\odot$ / (K \kms pc$^2$). (8) gas fraction (ratio of gas mass to stellar mass). (9) dust continuum}
\label{tab:CO_props}
\end{table*}
%M$_\odot$ / (K \kms pc$^2$)n,

\twocolumn

\begin{figure*}
\centering
\includegraphics[width=0.71\textwidth]{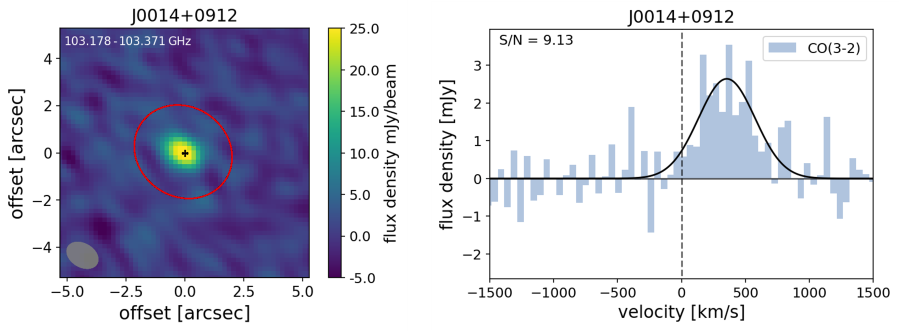}
\includegraphics[width=0.71\textwidth]{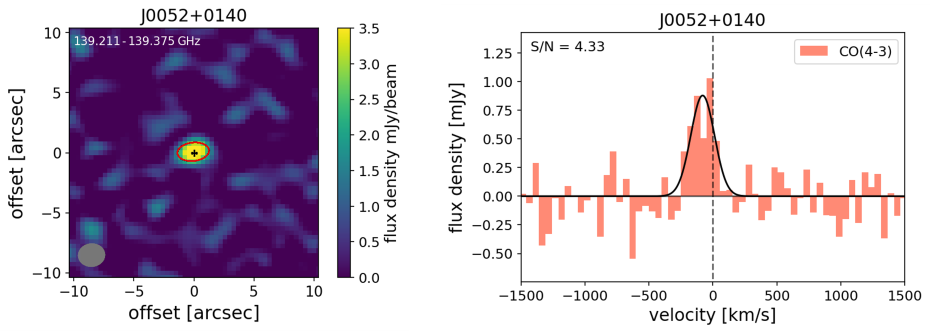}
\includegraphics[width=0.71\textwidth]{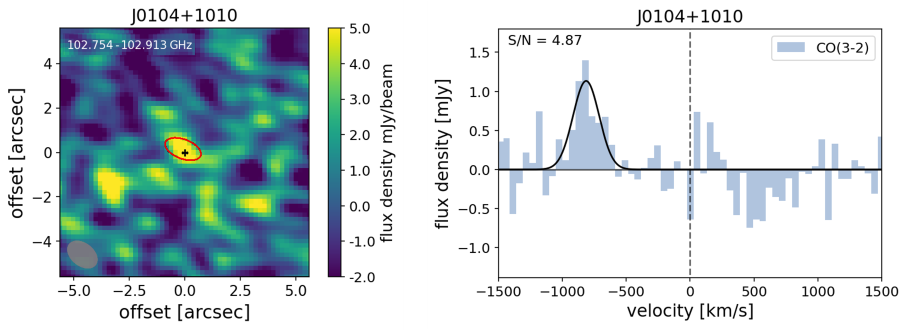}
\includegraphics[width=0.71\textwidth]{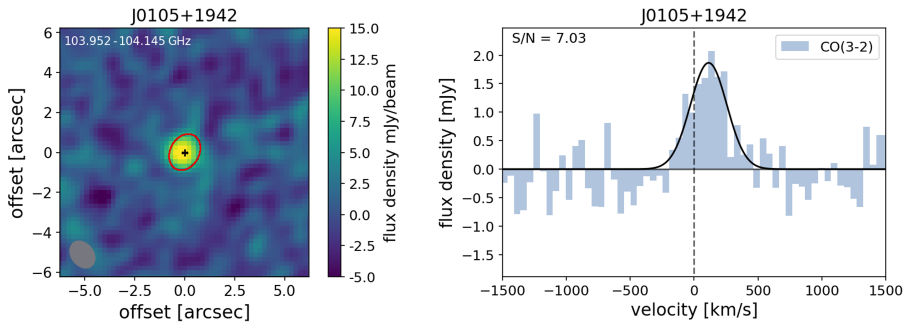}
\includegraphics[width=0.71\textwidth]{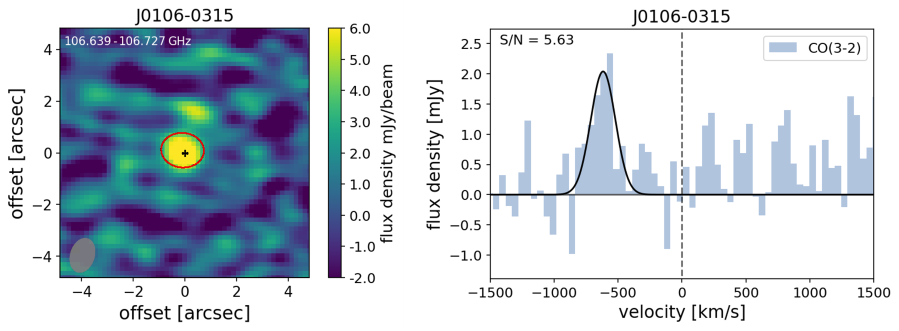}
\caption{Left panel: Narrowband image collapsed over the frequency range indicated in the top left of the cutout. The black cross indicates the centre of the observation. The grey ellipse represents the corresponding beam size. The red ellipse indicates the region where the spectrum is extracted from. The flux density in mJy/beam is indicated by the colourbar. Right panel: Spectrum extracted from the region indicated by the red ellipse in the left panel. 
}
\label{fig:all_spectra}
\end{figure*}

\begin{figure*}
\ContinuedFloat 
\centering
\includegraphics[width=0.71\textwidth]{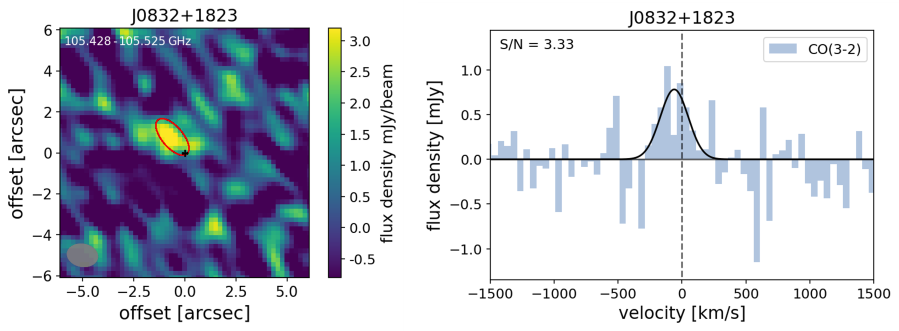}
\includegraphics[width=0.71\textwidth]{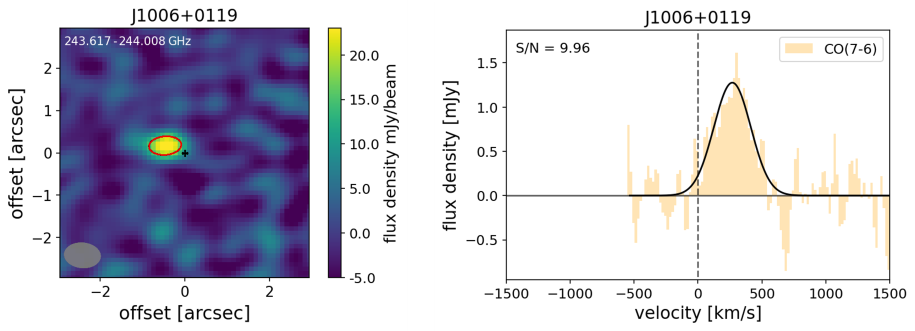}
\includegraphics[width=0.71\textwidth]{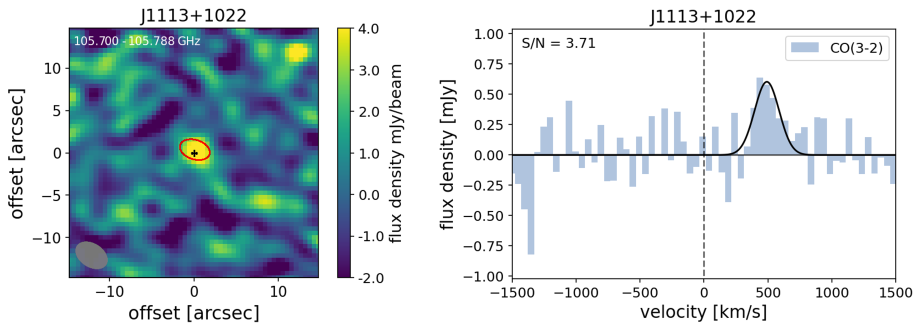}
\includegraphics[width=0.71\textwidth]{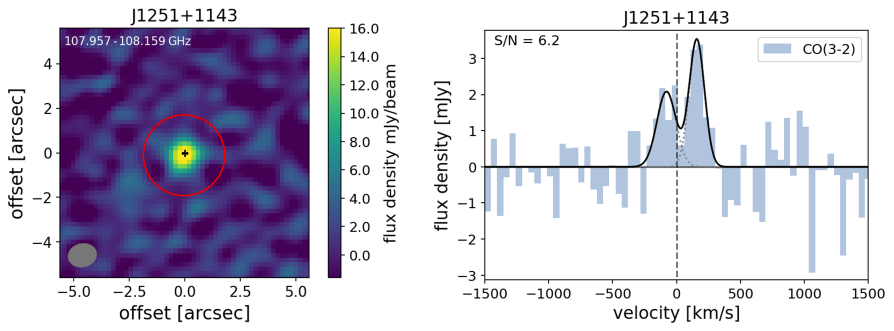}
\includegraphics[width=0.71\textwidth]{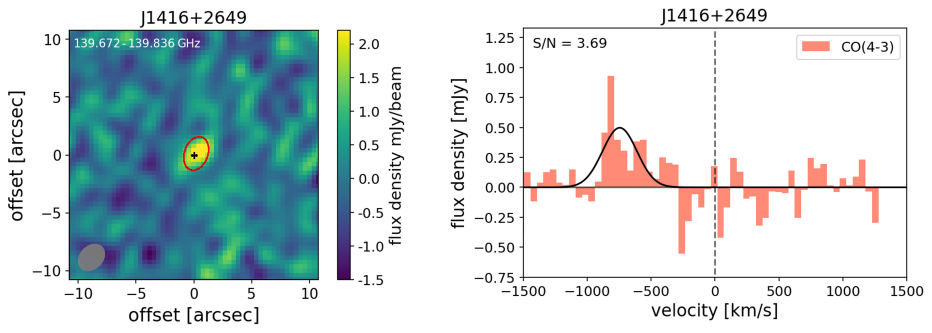}

\caption{continued.}
\end{figure*}

\begin{figure*}
\ContinuedFloat 
\centering
\includegraphics[width=0.71\textwidth]{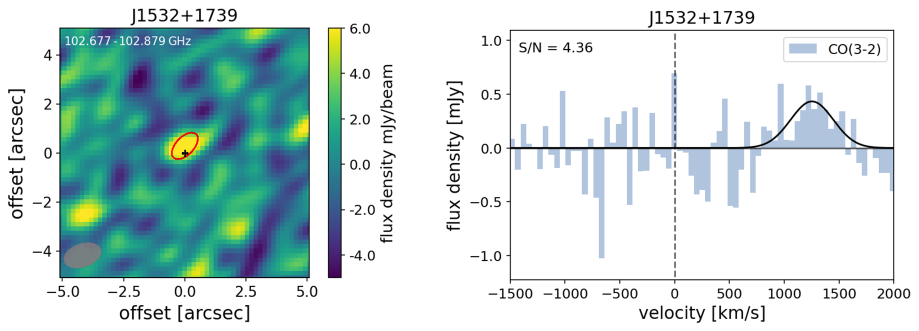}
\includegraphics[width=0.71\textwidth]{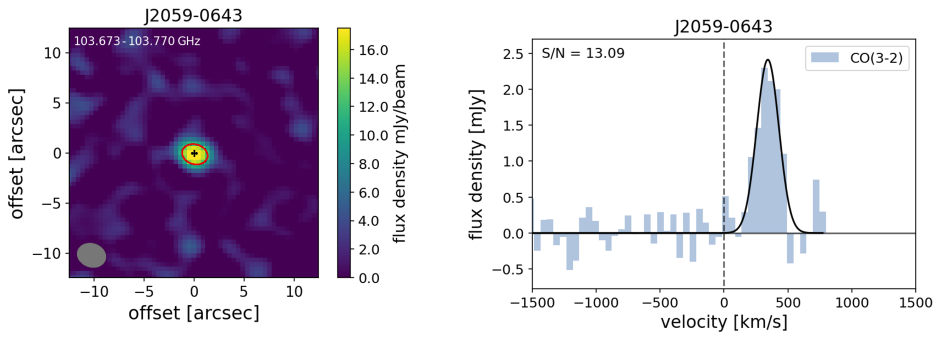}
\includegraphics[width=0.71\textwidth]{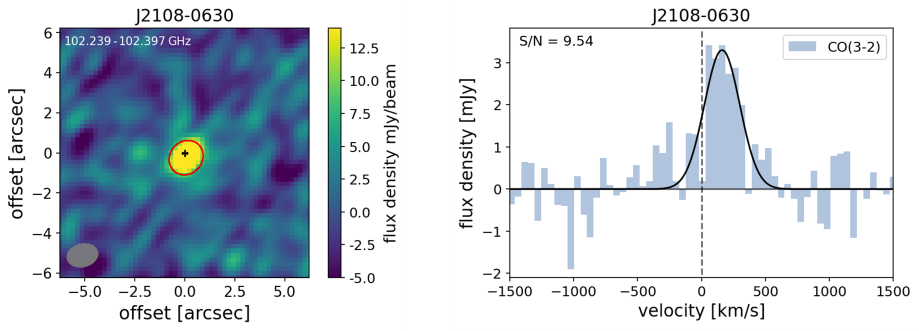}
\includegraphics[width=0.71\textwidth]{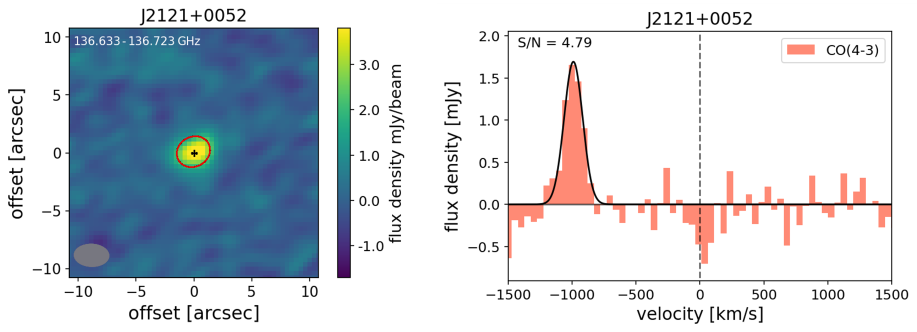}
\includegraphics[width=0.71\textwidth]{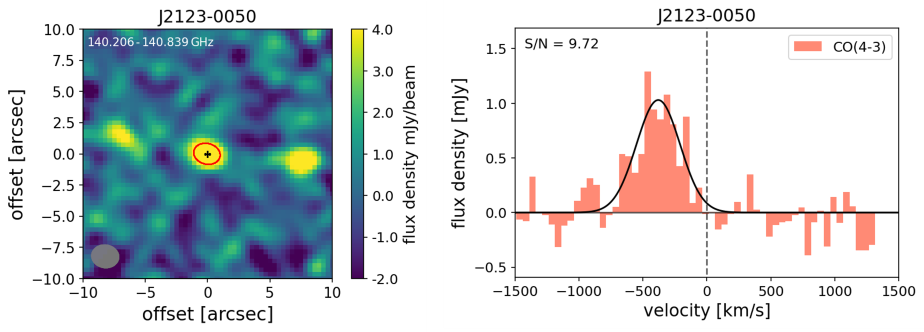}

\caption{continued.}
\end{figure*}

\begin{figure*}
\ContinuedFloat 
\centering
\includegraphics[width=0.71\textwidth]{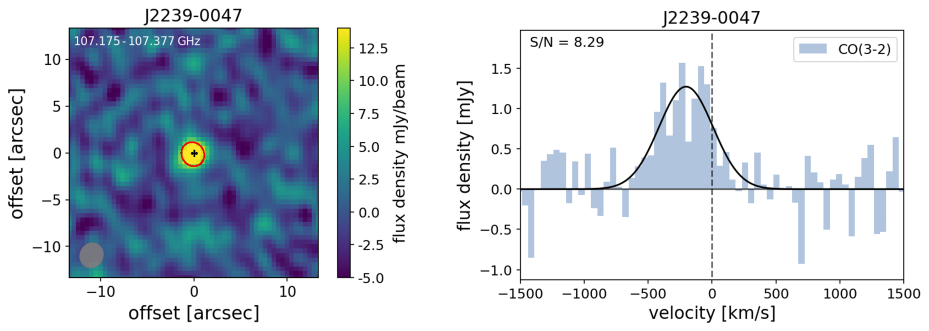}
\includegraphics[width=0.71\textwidth]{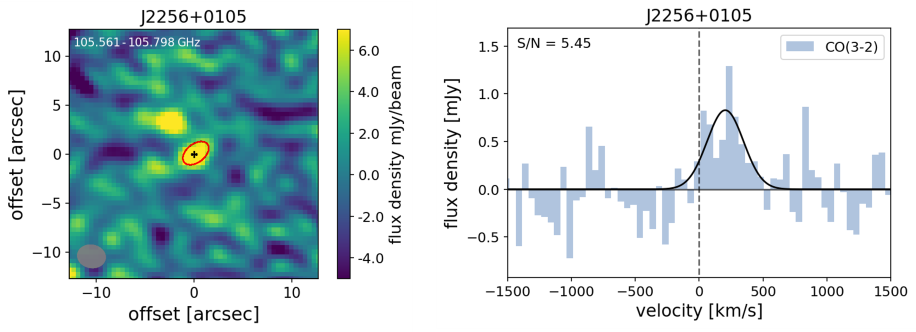}
\includegraphics[width=0.71\textwidth]{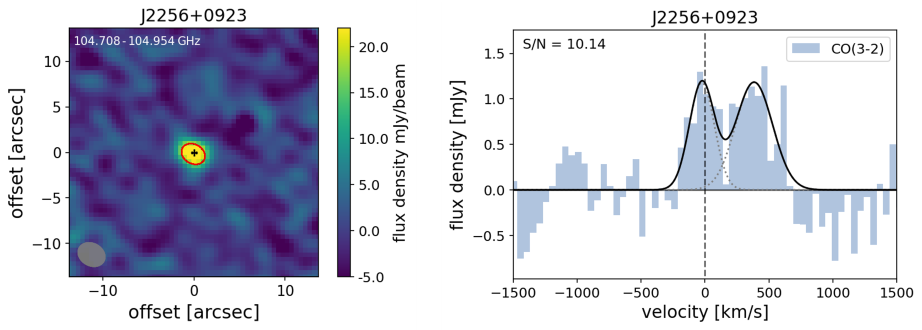}
\includegraphics[width=0.71\textwidth]{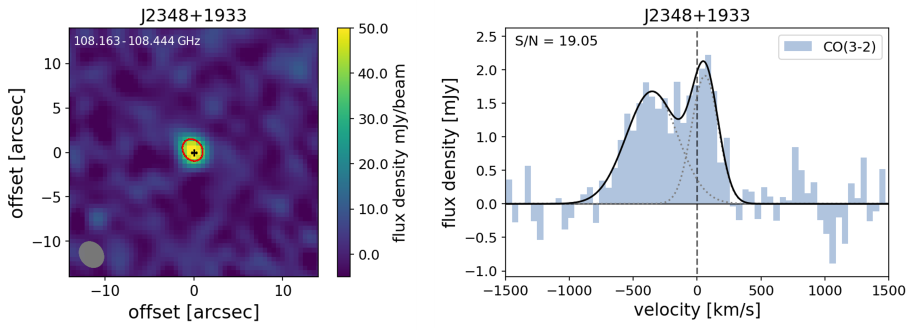}
\includegraphics[width=0.71\textwidth]{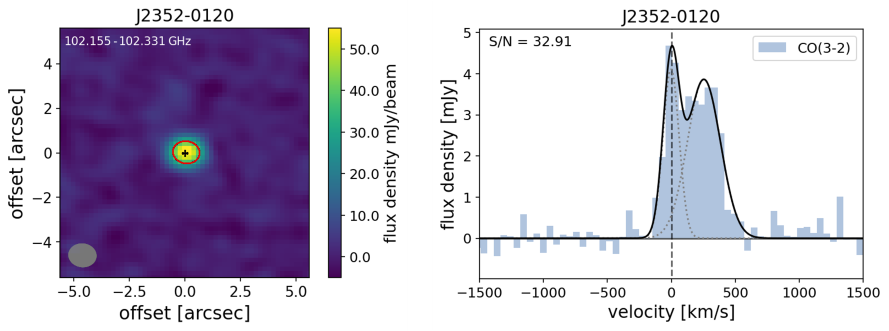}
\caption{continued.}
\end{figure*}

\begin{figure*}
\centering
\includegraphics[width=0.71\textwidth]
{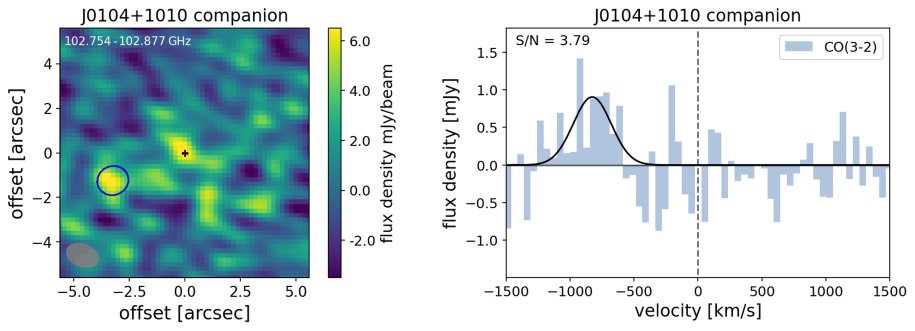}
\includegraphics[width=0.71\textwidth]{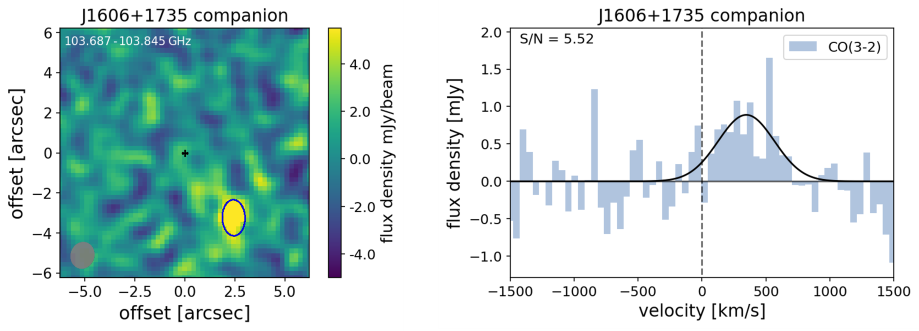}
\includegraphics[width=0.71\textwidth]{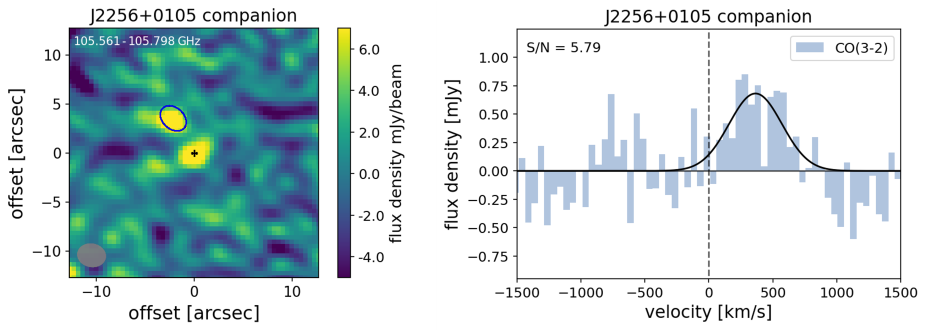}
\includegraphics[width=0.71\textwidth]{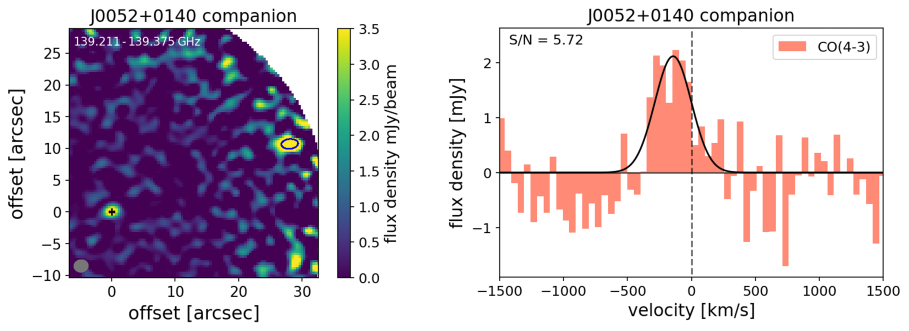}
\caption{Spectra of companion galaxies to the quasars. Left panel: Narrowband image collapsed over the frequency range indicated in the top left of the cutout. The black cross indicates the centre of the observation. The grey ellipse represents the corresponding beam size. The blue ellipse indicates the companion galaxy from which the spectrum is extracted. The flux density in mJy/beam is indicated by the colourbar. Right panel: spectrum extracted from source around the region indicated by the red ellipse in the left panel. 
}
\label{fig:companion_spectra}
\end{figure*}

\begin{figure*}
\ContinuedFloat 
\centering
\includegraphics[width=0.71\textwidth]{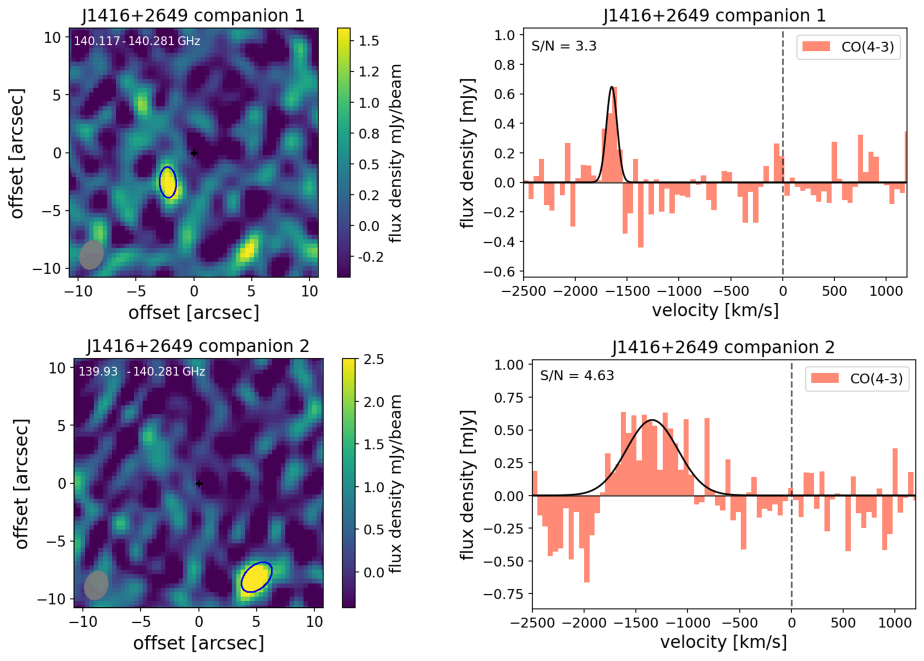}
\includegraphics[width=0.71\textwidth]{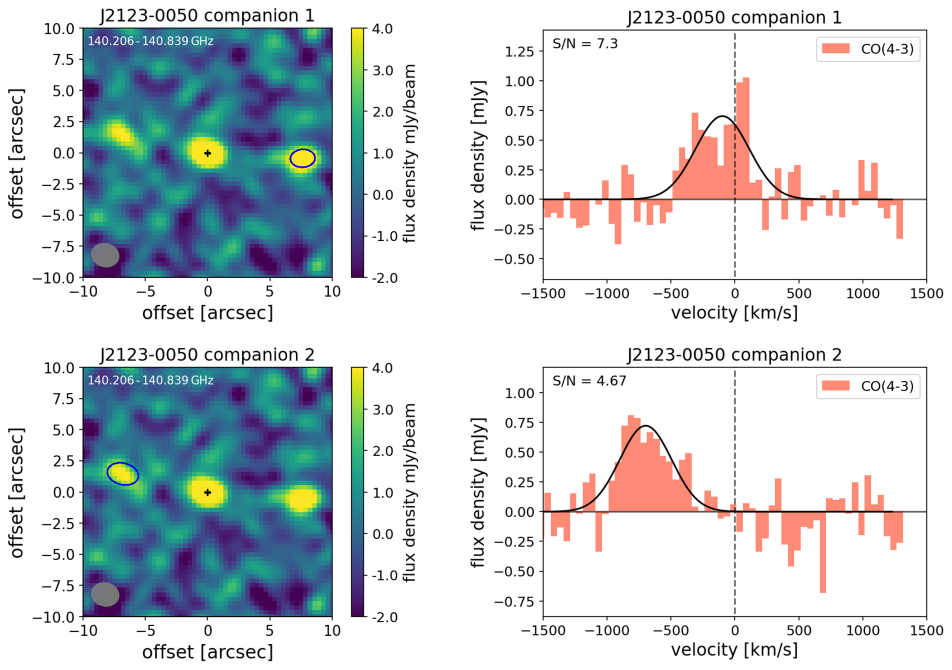}
\caption{continued.}
\end{figure*}

\begin{figure*}
\centering
\includegraphics[width=0.32\textwidth]{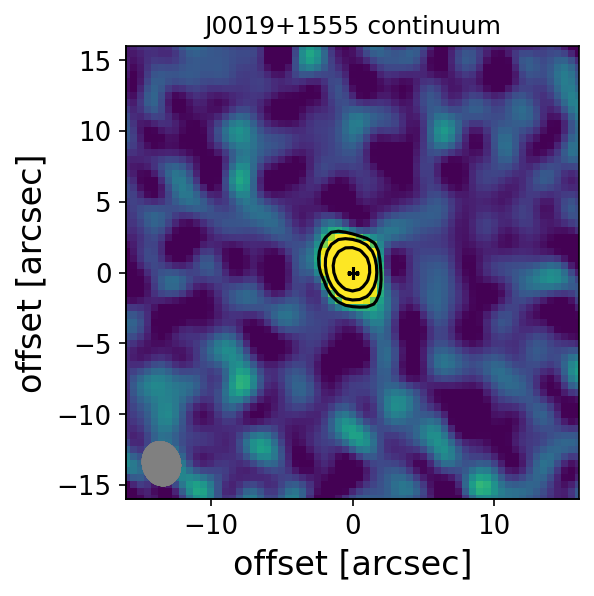}
\includegraphics[width=0.32\textwidth]{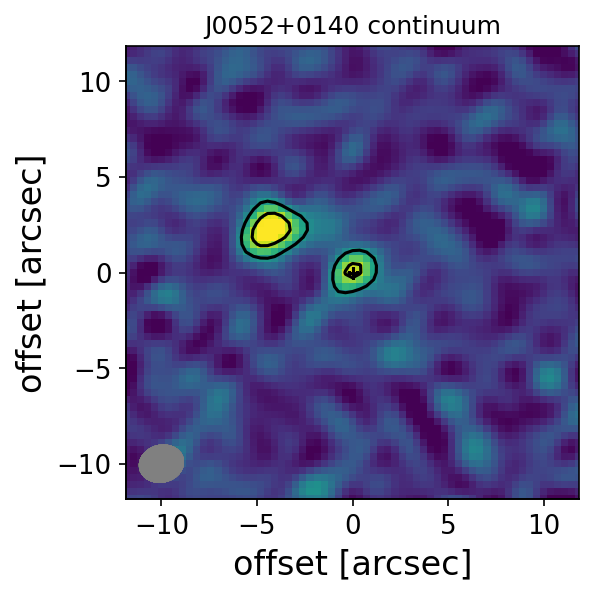}
\includegraphics[width=0.32\textwidth]{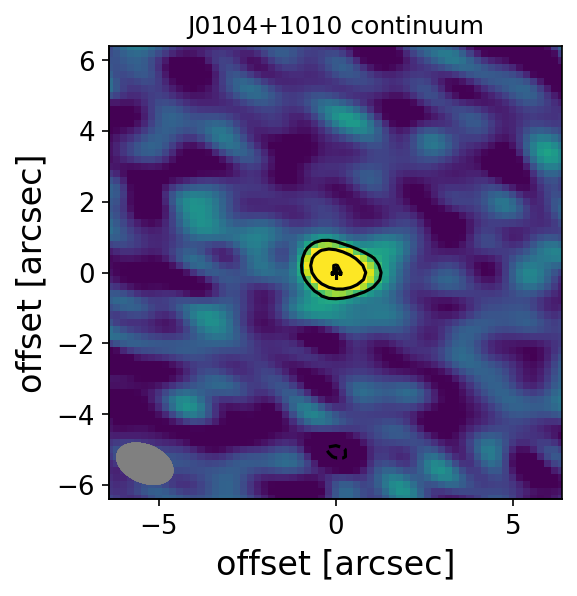}
\includegraphics[width=0.32\textwidth]{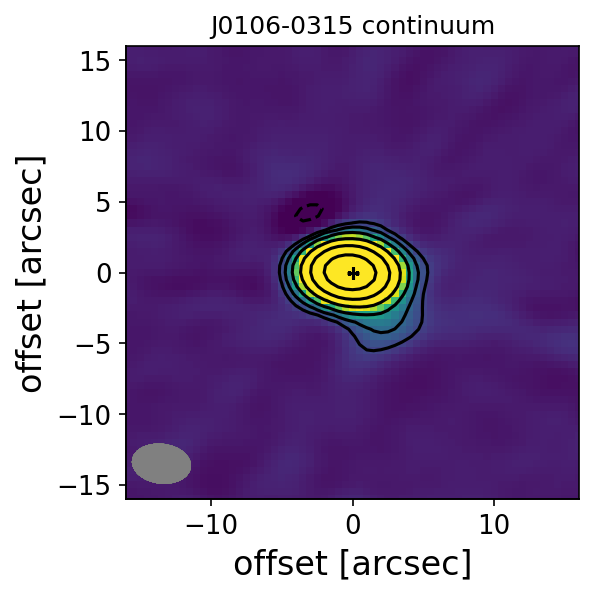}
\includegraphics[width=0.32\textwidth]{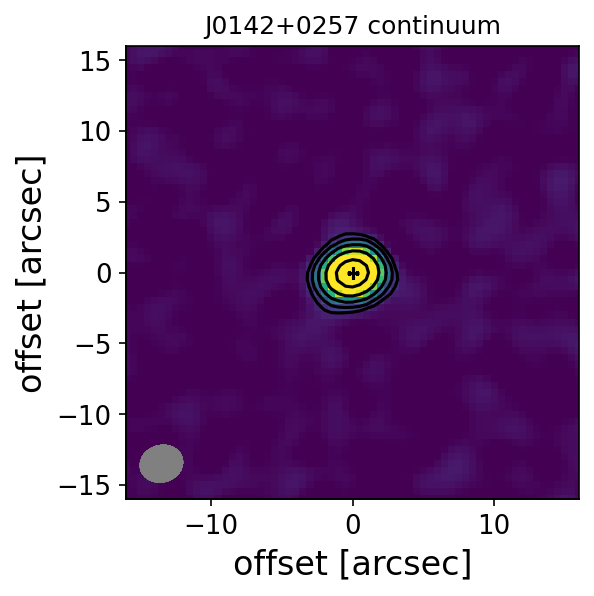}
\includegraphics[width=0.32\textwidth]{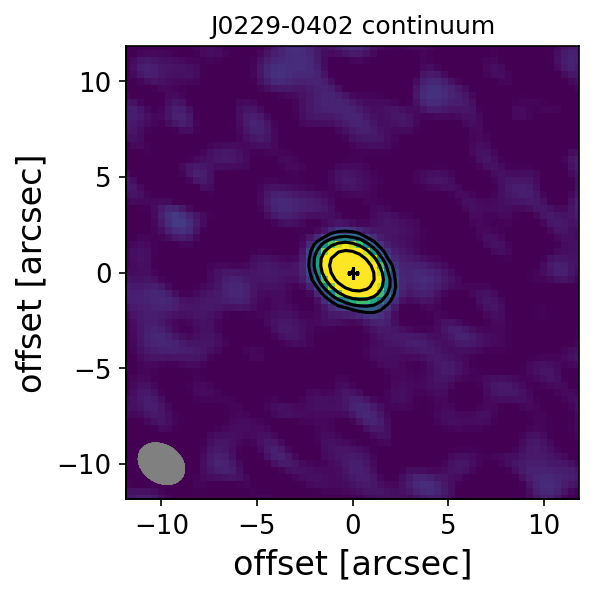}
\includegraphics[width=0.32\textwidth]{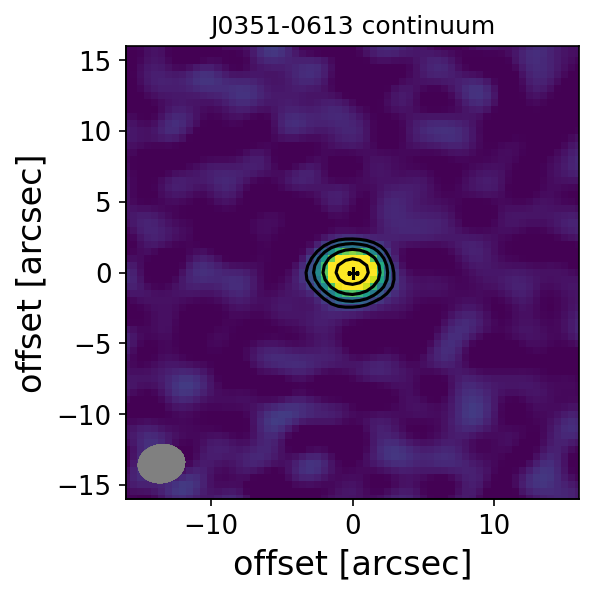}
\includegraphics[width=0.32\textwidth]{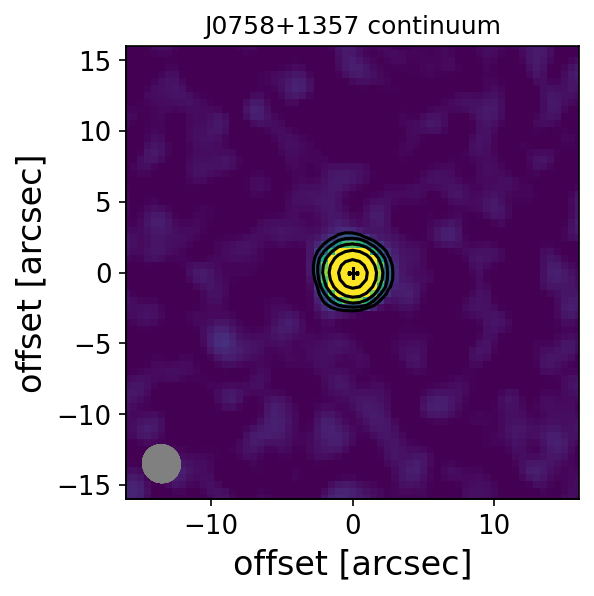}
\includegraphics[width=0.32\textwidth]{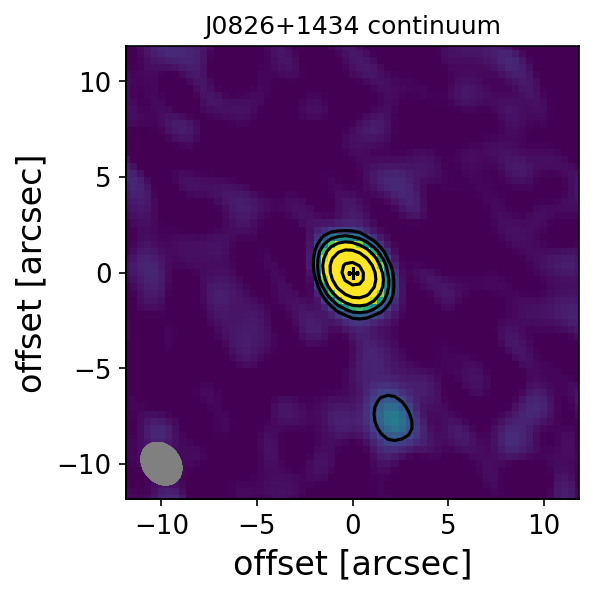}
\includegraphics[width=0.32\textwidth]{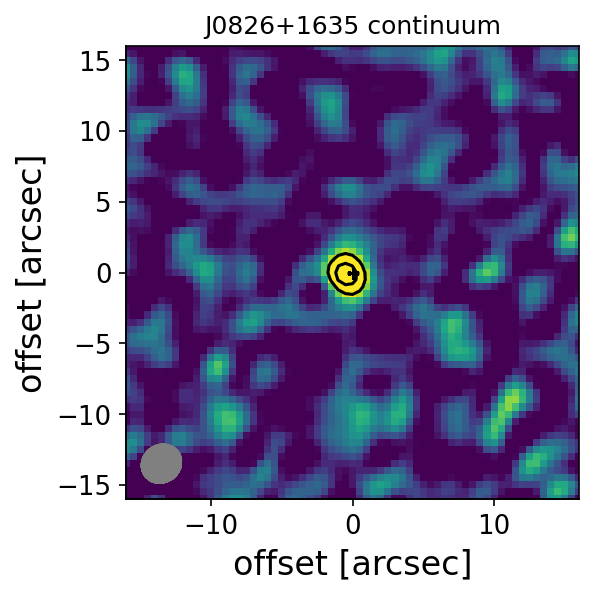}
\includegraphics[width=0.32\textwidth]{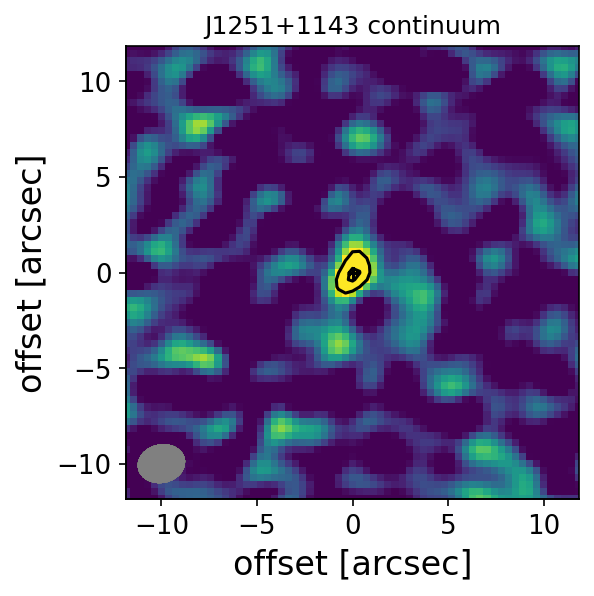}
\includegraphics[width=0.32\textwidth]{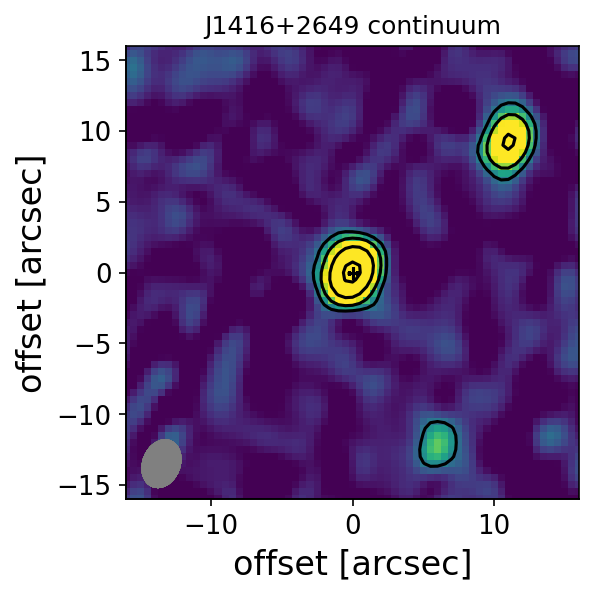}
\caption{Continuum images for all those with continuum detections. Black crosses indicate the centre of observations. Black contours begin at 2 sigma. Black dashed contours correspond to negative 2 sigma (where present).
}
\label{fig:cont_cutouts}
\end{figure*}
\begin{figure*}
\ContinuedFloat 
\centering
\includegraphics[width=0.32\textwidth]{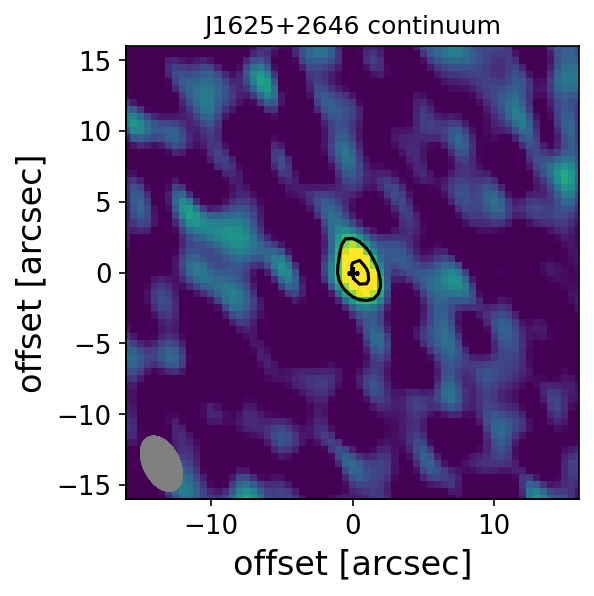}
\includegraphics[width=0.32\textwidth]{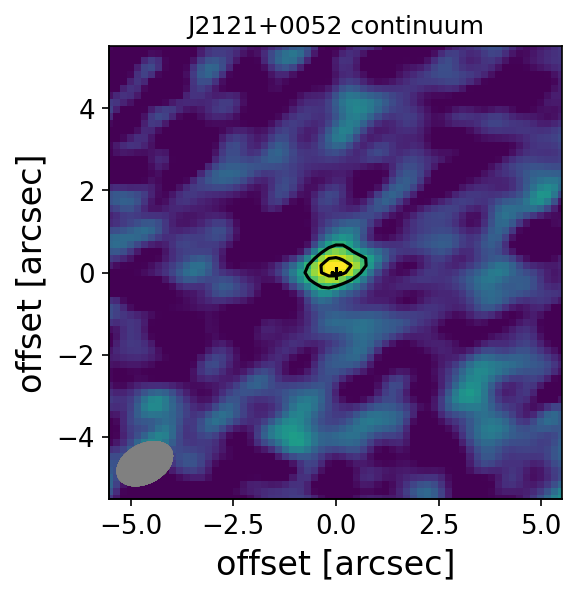}
\includegraphics[width=0.32\textwidth]{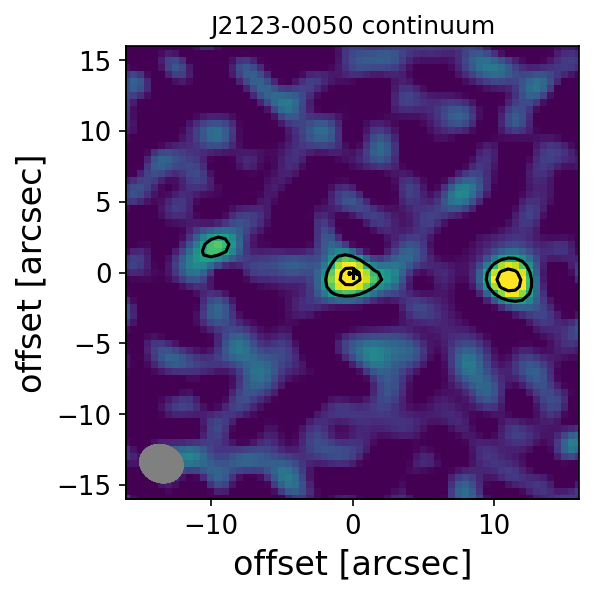}
\includegraphics[width=0.32\textwidth]{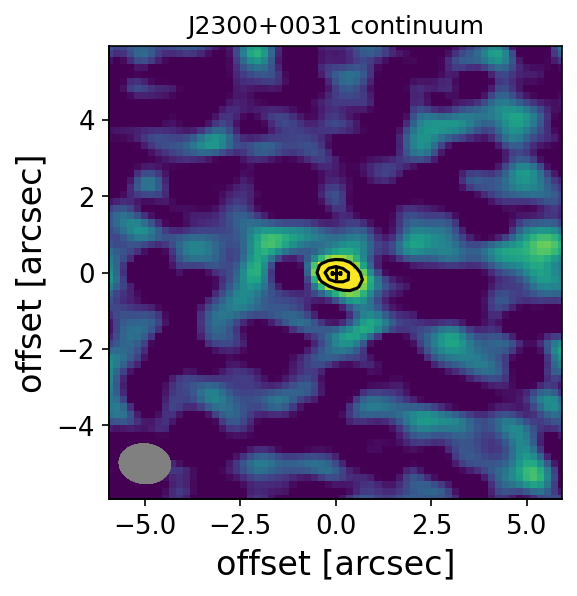}
\includegraphics[width=0.32\textwidth]{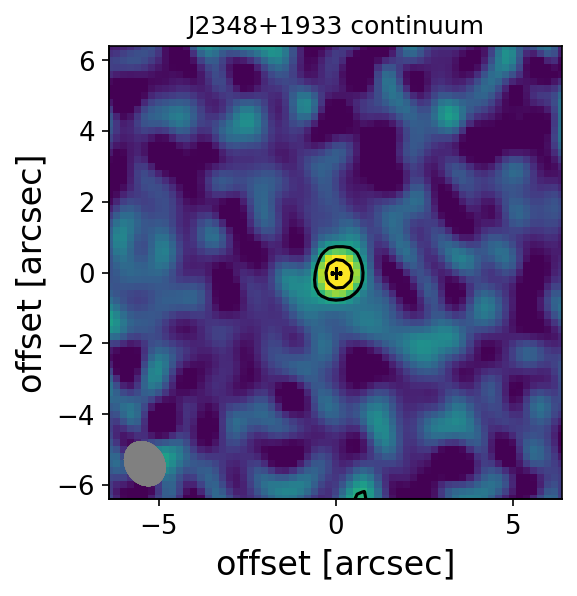}
\includegraphics[width=0.32\textwidth]{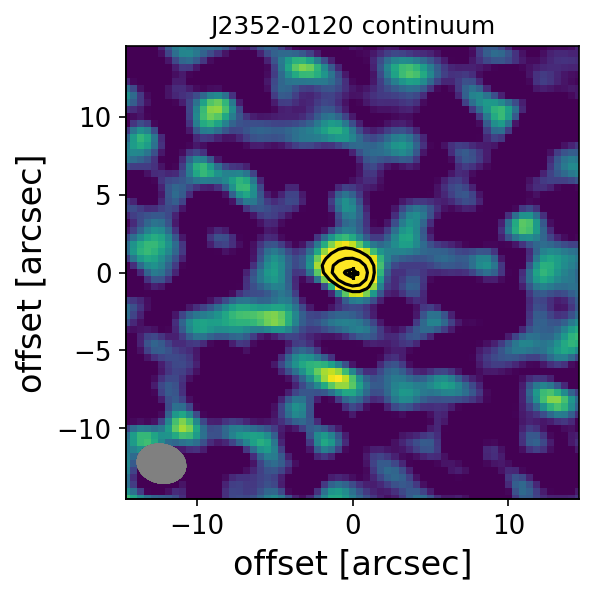}
\caption{continued.}
\end{figure*}

\bibliographystyle{mnras}
\bibliography{Molyneux+24} % if your bibtex file is called example.bib

%\begin{thebibliography}{99}
%\end{thebibliography}

%%%%%%%%%%%%%%%%%%%%%%%%%%%%%%%%%%%%%%%%%%%%%%%%%%

%%%%%%%%%%%%%%%%% APPENDICES %%%%%%%%%%%%%%%%%%%%%

%%%%%%%%%%%%%%%%%%%%%%%%%%%%%%%%%%%%%%%%%%%%%%%%%%

% Don't change these lines
\bsp	% typesetting comment
\label{lastpage}
\end{document}